\documentclass[epj]{svjour}
\usepackage[pdftex]{graphics}
\begin{document}
\title{The equation of state and symmetry energy of low density nuclear matter}
\author{K.~Hagel\inst{1} \and J.~B.~Natowitz\inst{1} \and G.~R\"{o}pke\inst{2}
}                     
%
%
\institute{Cyclotron Institute, Texas A \& M University, College Station, 
Texas 77843-3366, USA \and Institut f\"{u}r Physik, Universit\"{a}t Rostock,
Universit\"{a}tsplatz 3, D-18055 Rostock, Germany}
\date{Received: date / Revised version: date}
%
\abstract{
The symmetry energy of nuclear matter is a fundamental ingredient in
the investigation of exotic nuclei, heavy-ion collisions and astrophysical 
phenomena.  A recently developed quantum statistical (QS) approach 
that takes the formation of clusters into account predicts low density 
symmetry energies far above the usually quoted mean field limits. A 
consistent description of the symmetry energy has been developed that 
joins the correct low-density limit with values calculated from 
quasi-particle approaches valid near the saturation density. The results 
are confronted with experimental values for free symmetry energies and 
internal symmetry energies, determined at sub-saturation densities and 
temperatures below 10~MeV using data from heavy-ion collisions. There 
is very good agreement between the experimental symmetry energy values 
and those calculated in the QS approach %
} 
\maketitle
\section{Introduction}
\label{intro}

Reliable understanding of the density dependence of the nuclear equation of 
state (EOS) over a wide range of densities and temperatures is a very 
important need in the investigation of both nuclear and astrophysical 
phenomena. One fundamental ingredient of the EOS which is the subject of many 
recent experimental and theoretical investigations is the symmetry 
energy $E_{\rm{sym}}(n,T)$ that describes the dependence of the energy per 
nucleon on the proton to neutron ratio.  It governs phenomena from nuclear 
structure to astrophysical processes.  As a result, the variation of the 
symmetry energy with nucleon density $n=n_n+ n_p$ with $n_n, n_p$ denoting 
the neutron and proton density, respectively, and temperature $T$ has been 
extensively investigated in recent years.  A recent review is given by 
Li~\textit{et al.}~\cite{Li:2008gp}; see also~\cite{Lattimer:2006xb} and 
other contributions to this volume of the European Physical Journal A, in 
particular~\cite{Typel2013}. 

\begin{figure*}
\resizebox{1.2\textwidth}{!} { %
\hspace {-3.5truein}
  \includegraphics{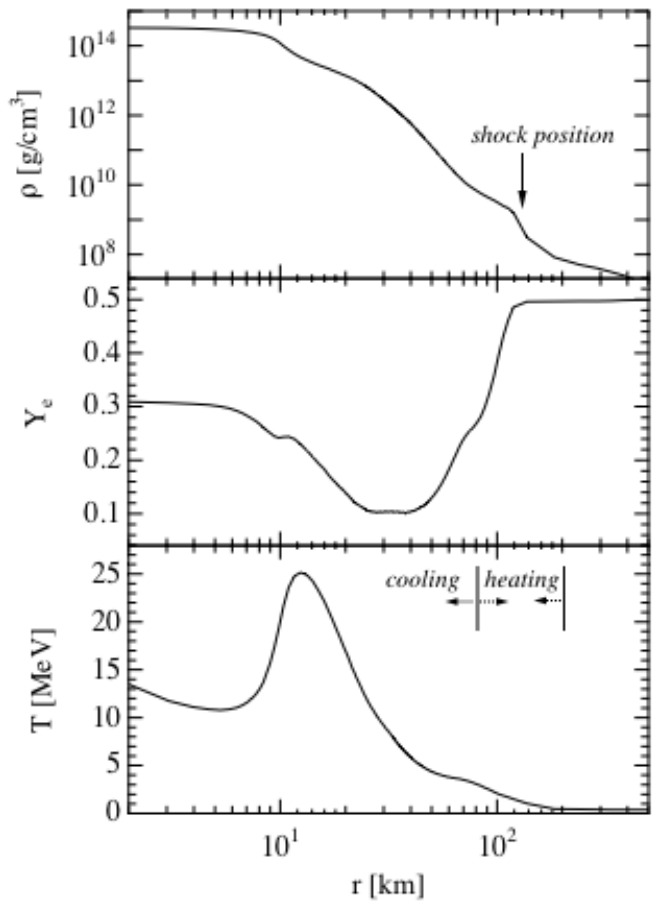}
\hspace {-5.5truein}
  \includegraphics{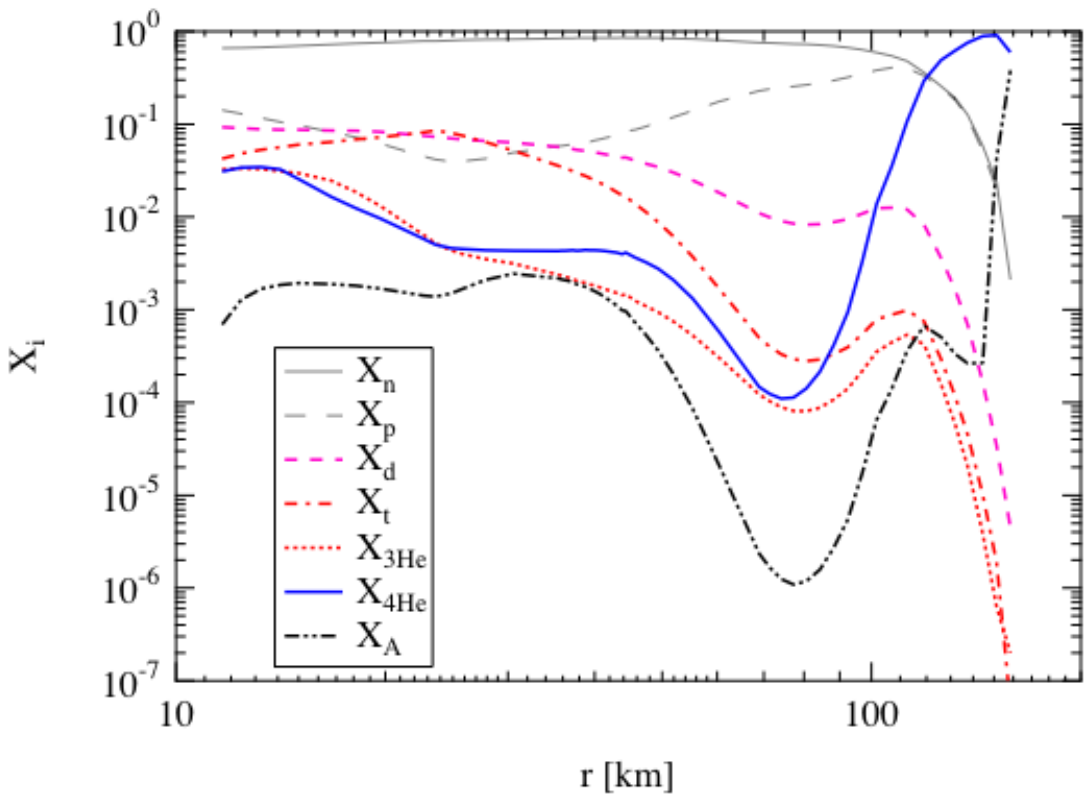}
}
\caption{Left, density, electron fraction and temperature radial profiles 
of a post core bounce supernova. Right, mass fractions for $Z=1$ and 2 
nuclei as well as  $A > 4$ nuclei (metals) as a function of radial distance.
See Ref.~\cite{Sumiyoshi:2008}.
}
\label{fig:1}       
\end{figure*}

While conventional mean-field approaches to the EOS treat nuclear matter 
as uniform, it is well established that the properties of low density 
nuclear matter are governed by correlations, in particular by the 
appearance of bound states, i.e. clusters.  That correlations and 
clusterization are important and exhibit significant density dependences 
may be seen, for example, in Figure~\ref{fig:1}~\cite{Sumiyoshi:2008}.  
In the left side of that figure the calculated density, electron fraction 
and temperature distributions of a post core collapse supernova are 
presented as a function of radial distance. At the larger radii 
temperatures and densities which are accessible in near Fermi energy 
collisions~\cite{Kowalski:2006ju},  are seen. In the right 
side of the picture are particle mass fractions calculated by 
Sumiyoshi \textit{et al.}~\cite{Sumiyoshi:2008} for $Z=1$ and $Z=2$ nuclei 
as a function of radial distance. Furthermore, also the mass fraction of 
higher mass nuclei (metals) are shown, that is calculated as the sum over 
all elements with $A > 2$ . The role of the cluster formation in the 
neutrino-sphere region (the region of last neutrino interaction) is of 
particular interest~\cite{OÕConnor,lentz12}.  As will be seen below, 
this region of temperature and density is accessible in collisions of 
heavy ions at intermediate energies. 
       
    Currently some extensive EOS calculations and existing tabulations, 
based on varying nucleon-nucleon interactions, serve as standard input for 
a wide variety of astrophysical simulations.  In general these calculations 
involve a number of simplifying assumptions and allow for a limited number 
of clustered species at lower densities where important clustering effects are 
expected~\cite{Lattimer:1991nc,Shen:1998gq,Shen:2011,Hempel:2010,Horowitz:2005nd}.
We show that new experiments allow for a laboratory test of the EOS 
and the symmetry energy in the low-density region.
The results are in accordance with a systematic QS approach that takes 
correlations and cluster formation in warm dense matter into account.

\section{Clusterization and the symmetry  energy at very low density}
\subsection{Nuclear binding energies and Bethe-Weizs\"{a}cker mass formula}
The symmetry energy characterizes the dependence of the nuclear binding 
energy on the asymmetry 
\begin{equation}
\label{asymm}
\delta=(N-Z)/A=1-2Y_p
\end{equation}
with $Z$ and $N$
the proton and neutron numbers, and $A=N+Z$. In astrophysics, the asymmetry 
is usually specified by the total proton fraction $Y_p$.  For charge neutral 
stellar matter $Y_p$ is equal to the electron fraction $Y_e$.  The definition 
of the symmetry energy can be expressed in different ways, 
see~\cite{Typel2013}.  Although in the binding energy of nuclei, there is 
a dependence on $\delta$ resulting from the Coulomb interaction, the 
symmetry energy does not include this trivial effect. Rather, it is defined 
by the asymmetry contribution of the strong interaction to the binding energy,
as expressed within the  Bethe-Weizs\"{a}cker mass formula 
\begin{eqnarray}
\label{eq:BW}
 B(N,Z) & = &  a_{V} A - a_{S} A^{2/3} 
 \\ \nonumber & & 
- \left( a_{V}^{(\rm sym)} A + a_{S}^{(\rm sym)} A^{2/3} \right)
 \delta^{2} 
 \\ \nonumber & & 
 - a_{C} \frac{Z(Z-1)}{A^{1/3}} + \dots 
\end{eqnarray}
where volume, surface, volume symmetry, surface symmetry and Coulomb
contributions exhibit  a particular dependence on the mass number
$A$ and asymmetry $\delta$. (Pairing and other features such as shell
effects are not considered here.)  The parameters $a_{V}$, $a_{S}$, 
$a_{V}^{(\rm sym)}$, $a_{S}^{(\rm sym)}$, $a_{C}$ are generally found
by fitting nuclear masses
\begin{equation}
m_{N,Z} = N m_{n} + Z m_{p} - B(N,Z)
\end{equation}
across the whole chart of known nuclei. Here $m_{n}$ and $m_{p}$ are the
neutron and proton rest masses, respectively. Typical values of the
coefficients $a_{V},a_{S},a_{V}^{(\rm sym)}, a_{S}^{(\rm sym)},a_{C}$ are given in Ref.~\cite{Lun03}.

Our empirical knowledge of the symmetry energy near the saturation density, 
$n_{0}$, is based primarily on the binding energy analysis.  The 
Bethe-Weizs\"{a}cker mass formula leads to values of about 
$E_{\rm sym}(n_0,0)=28-34$ MeV for the nuclear matter symmetry energy at 
zero temperature and saturation density $n_{0} \approx 0.16$ fm$^{-3}$, 
if surface asymmetry effects are properly taken into account~\cite{Dan03}.
A recent analysis of experimental data~\cite{jiang12} gives the value 
$a_{V}^{(\rm sym)} = 32.1  \pm 0.3$ MeV.

Note that in the above, the separation of the Coulomb energy has been  
performed under the assumption that the strong interaction part is 
symmetric with respect to $\delta \to - \delta$ so that the symmetry energy 
contains no linear term in $\delta$. Then, the Coulomb term in 
Eq.~(\ref{eq:BW}) contains besides the electrostatic field energy also the 
contribution of the nuclear wave function that may change with the 
asymmetry $\delta$.

\subsection{Symmetry energy of nuclear matter at finite temperature}

Within a more fundamental approach, we can consider a given state of the 
nuclear system described by the wave function ($| \Psi \rangle$) or the 
statistical operator ($\rho$) and calculate the expectation value of the 
energy.  In particular, we consider homogeneous matter at finite 
temperature $T$ consisting of nucleons (neutrons $n$, protons $p$) and 
electrons ($e$).  To eliminate the dependence on the volume $V$ of the 
system that goes to infinity in the thermodynamic limit, we introduce the 
neutron density $n_n=N_n/V$ and the proton density $n_p=N_p/V$ that remain 
finite.  The electron density $n_e=n_p$ is necessary to make the system 
globally charge neutral, to avoid diverging Coulomb energy. It is determined 
by the proton density and cannot be considered as an additional degree of 
freedom.  Note that we neglect weak interaction processes that lead 
to $\beta$ equilibrium so that in full thermodynamic equilibrium the 
asymmetry  (that follows from Eq. (\ref{asymm}) after dividing the nucleon 
numbers by the volume $V$)
\begin{equation}
\label{asymm1}
\delta=\frac{n_n-n_p}{n},\qquad n=n_n+n_p
\end{equation}
is an independent thermodynamic parameter, in addition to $T$ and the total 
baryon density $n$.  We note that the inclusion of further elementary 
particles such as muons, hyperons or neutrinos may be of interest in 
astrophysical context, at higher densities and temperatures.

For this "frozen" hot matter characterized by $T, n, \delta$ where weak 
interaction processes are neglected, thermodynamic equilibrium corresponds 
to the minimum of the free energy
\begin{equation}
\label{freeen}
{\tilde F}(T,V,N_n, N_p)=V f(T, n, \delta)=N F(T, n, \delta), 
\end{equation}
$N=n V$.  To eliminate the dependence on the volume $V$ of the system that 
goes to infinity, we introduce the density of the free energy 
$f(T, n, \delta)$ or the free energy per nucleon 
$F(T, n, \delta)= f(T, n, \delta)/n$.  Being a thermodynamic potential, all 
other thermodynamic quantities such as pressure, entropy or internal energy
are consistently derived using the thermodynamic relations.

It is convenient to describe the thermodynamic equilibrium by the grand 
canonical ensemble that is given by a Gibbs distribution 
\begin{equation}
\label{grandcan}
\rho_{\rm gr. can.}(T,\mu_n,\mu_p)=\frac{1}{Z(T,\mu_n,\mu_p)} {\rm e}^{-(H-\mu_n N_n-\mu_p N_p)/T}
\end{equation}
with the neutron chemical potential $\mu_n$ and the the proton chemical 
potential $\mu_p$, $Z(T,\mu_n,\mu_p)$ is the partition function.  The 
chemical potential of the background charge compensating electrons is not a 
new degree of freedom but is fixed by the condition of charge neutrality.
As long as there are no phase transitions, the various statistical ensembles 
are equivalent and give identical results for the equation of state.  The 
chemical potentials are related to the particle densities by the following 
equations of state
\begin{eqnarray}
\label{density}
n_n(T, \mu_n,\mu_p)&=&{\rm Tr} \{ \rho_{\rm gr. can.}(T,\mu_n,\mu_p) {\hat N}_n\}/V ,\nonumber \\
n_p(T, \mu_n,\mu_p)&=&{\rm Tr} \{ \rho_{\rm gr. can.}(T,\mu_n,\mu_p) {\hat N}_p\}/V\,.
\end{eqnarray}
In momentum representation, the particle number operators are $ {\hat N}_\tau=\sum_{{\bf p},\sigma} a^\dagger_ka^{}_k$ where the quantum number $k$ 
contains besides momentum $\bf p$ and spin $\sigma$ also the isospin $\tau$.  
We use these thermodynamic relations later on to calculate the free energy 
by integrating the Helmholtz equation, see Eq. (10) of 
Ref.~\cite{Typel:2009sy} and Eq. (\ref{freV}) below.

The Hamiltonian $H$ contains the kinetic energy (we use relativistic 
kinematics), the strong interaction and the Coulomb interaction that reads 
in position representation
\begin{equation}
\label{Coulomb}
{\hat V}^{\rm Coul}=\frac{1}{2}\sum_{c,d}\int d^3 r \psi_c^\dagger({\bf r})\psi_d^\dagger({\bf r}')\frac{e_ce_d}{|{\bf r}-{\bf r}'|}\psi_d({\bf r}')\psi_c({\bf r})
\end{equation}
where the index $c,d$ denotes the different components, including spin.
Calculating the expectation value of total energy 
\begin{eqnarray}
\label{energy}
&&U^{\rm total}(T, V,\mu_n,\mu_p)={\rm Tr} \{ H \rho_{\rm gr. can.}(T,\mu_n,\mu_p)\} \\
&& = U(T, V,\mu_n,\mu_p)+ U^{\rm Coul}(T, V,\mu_n,\mu_p)+U^{\rm el}(T, V,\mu_n,\mu_p)\nonumber
\end{eqnarray}
for this equilibrium quantum state, one can 
separate the contribution of the Coulomb energy 
\begin{equation}
\label{Coulomb1}
 U^{\rm Coul}={\rm Tr} \{ \rho_{\rm gr. can.}(T,\mu_n,\mu_p) {\hat V}^{\rm Coul}\}
\end{equation}
as well as the contribution of the electron kinetic energy $U^{\rm el}$.
The symmetry energy follows from the expansion of the remaining energy (nucleon kinetic energy and strong interaction)
with respect to the asymmetry. 

Using the equations of state (\ref{density}) 
$n_n = n_n(T,\mu_n,\mu_p)$, $n_p = n_p(T,\mu_n,\mu_p)$ we replace the 
thermodynamic variables $\mu_n,\mu_p$ by the densities $n_n, n_p$ or 
equivalently, the total baryon density $n$ and the asymmetry $\delta$. 
To obtain the symmetry energy per nucleon we consider the energy per 
nucleon $E(n,\delta,T)=U(T, V,\mu_n,\mu_p)/N$ (after subtracting the 
Coulomb energy according to Eq.~(\ref{energy})) as function of $\delta$.  
Because the strong interaction is assumed to be symmetric with respect to 
isospin, after subtracting the trivial term due to the difference of the 
neutron and proton mass in the relativistic kinematics, the first derivative 
at $\delta = 0$ is approximately zero, so that we have 
 \begin{equation}
\label{eq:esym_defAbl}
 E(n,\delta,T)\approx  E(n,0,T) +(m_n-m_p) \delta+ E^"_{\rm sym}(n,T) \delta^2 +{\cal O}(\delta^4).
 \end{equation}
Here, the symmetry energy is related to the second derivative with respect 
to the asymmetry $\delta$.  This definition is appropriate for experimental 
investigations where we explore nuclei near the valley of stability.  In 
experiments performed for nuclear systems, $\delta$ is relatively small.

Another representation of the symmetry energy coefficient is the definition
 \begin{equation}
\label{eq:esym_def}
  E_{\rm sym}(n,T) = \frac{E(n,1,T)
  + E(n,-1,T)}{2} - E(n,0,T) .
\end{equation}
This definition is consistent with the use of the second order derivative
only if the dependence on $\delta$ is purely quadratic.  In particular in 
the case of phase transition and at very low temperatures, this is not 
exactly the case, see the discussion below and Ref.~\cite{Typel2013} within 
this volume.

Comparing with the definition of the symmetry energy for finite nuclei 
given in the previous subsection, we give some comments.  The generalization 
from ground state nuclei to finite temperatures $T$ is straightforward 
within the scope of thermodynamics. 

An essential topic is the role of the nucleon distribution as expressed by the 
wave function or the statistical operator.  Instead of the equilibrium 
statistical operator $\rho_{\rm gr. can.}$, Eq.~(\ref{grandcan}), 
non-equilibrium distributions can also be considered to investigate the 
average energy.  Some confusion is connected with different assumptions 
regarding the quantum state of the system, {\it i. e.} the choice of the wave 
function ($| \Psi \rangle$) or the statistical operator ($\rho$).  Sometimes 
an antisymmetrized product ansatz that leads to the mean-field approximation 
is taken.  However, this ansatz, appropriate for noninteracting fermion 
systems, is not correct in describing the ground state or the 
thermodynamic equilibrium at finite $T$ for the interacting nuclear system 
because all correlations are neglected.  To treat real systems, for instance 
in astrophysics, we have to take into account the correlations that are 
also present at thermodynamic equilibrium and are determined by the full 
Hamiltonian $H$.

More delicate is the subdivision of the symmetry energy in Eq.~(\ref{eq:BW})
into a volume part and a surface part. For homogeneous systems, the 
density does not depend on the position in coordinate space.
This is also correct for correlations and bound state formation if the 
center-of-mass motion is taken into account.  Correlations and bound state 
formation are seen in the two-particle distribution function 
(structure factor).  After separation of the center-of-mass motion, 
the internal wave function of a cluster is usually approximated within
a local-density approach. Thomas-Fermi models, droplet models, and 
gradient expansions can be applied to estimate the binding energy of a 
cluster. Thus, the definition of the symmetry energy (\ref{eq:esym_defAbl}) 
is not in conflict with the Bethe-Weizs\"acker formula (\ref{eq:BW}) that 
follows if the distribution of clusters (nuclei) is given ("frozen out").  
Neglecting the  center-of-mass motion, the internal density distribution 
of the cluster can be treated within a gradient expansion that gives, 
within the local density approximation, the volume term and the surface term.
However, this subdivision, that is useful for some approximations, 
is not rigorous and cannot be applied, for instance, to describe the 
contribution of light clusters such as deuterons or $\alpha$ particles.

%

We also note that in the uncorrelated state of symmetric nuclear matter, 
where cluster formation is neglected (or the strong interaction is dropped), 
the Coulomb energy for charge neutral matter (proton and electron densities 
constant and compensating) does not vanish. Only the Hartree term is zero. 
Because of microfield fluctuations there is also a finite 
value of the Coulomb energy.  In the lowest approximation we obtain the Debye 
shift due to the two-particle distribution function (Montroll - World 
contribution to the virial expansion of the equation of state), see for instance Ref.~\cite{KKER}.

The Coulomb interaction occurs in two positions: the energy that is 
clearly additive and can separated, and the influence on the quantum state 
or structure that is given, for instance, by the equilibrium distribution 
or the ground state wave function. Without Coulomb interactions, the 
structure of nuclei would change significantly and would lead to arbitrarily 
large stable nuclei what is not realistic. Also the two-phase separation 
that is obtained considering only strong interaction is disfavored because 
of large Coulomb energy what leads to so-called pasta structures. 
(Note that due to the Hamiltonian that is used to describe the ground state 
or thermodynamic equilibrium, the Coulomb interaction also indirectly leads 
to a linear term $\propto \delta$ in the energy that can be compensated by 
a redefinition of the Coulomb term.)

\subsection{The symmetry energy at sub-saturation densities}
We introduce the symmetry energy of nuclear matter at finite temperature 
to characterize the properties of matter in astrophysics or in excited 
nuclei.  To describe real systems, the quantum statistical approach 
is used to introduce well-defined quantities.  Then, special approximations 
can be performed that may lead to different results, due to the 
approximations used.  Thus, different results for the symmetry energy of 
nuclear matter can be found in the literature that describe experiments 
with real systems to a good approximation only if the relevant processes 
are taken into account.  In the case of the symmetry energy of low density 
nuclear matter considered here, correlations and cluster formation are 
essential and have to be taken into account.  Mean-field approaches such as 
Skyrme interaction or relativistic mean field (RMF) approximation are not 
able to describe adequately the quantum state of matter at low densities 
because all correlations are neglected.  Most of recently used treatments do 
not give the correct low-density limit that is governed by cluster formation, 
as discussed in the following subsection.


Two different pictures are used to describe nuclear matter at 
sub-saturation densities. Near the saturation density, $n_0$,  the 
Fermi-liquid (single quasiparticle) approach is used to describe nuclear 
systems.  This picture is confined to that region where correlations, in 
particular cluster formation, are not relevant.  To give a number, 
the density should be larger than $n_0/3$.  An alternative picture for 
hot nuclear systems is nuclear statistical equilibrium (NSE) and related 
models that treat a noninteracting gas consisting of all possible bound 
states in chemical equilibrium (mass action law).  This chemical picture is 
confined to the low-density region (below $5 \times 10^{-4}$ fm$^{-3}$) 
where the interaction between the constituents can be neglected.

Many theoretical investigations have been performed to estimate the 
behavior of the symmetry energy as a function of $n$ and $T$ 
(Li \textit{et al.}~\cite{Li:2008gp}, see 
also~\cite{Fuchs:2005yn,Klahn:2006ir}).  Typically, quasi-particle 
approaches such as Skyrme Hartree-Fock and relativistic mean field (RMF) 
models or Dirac-Brueckner Hartree-Fock calculations are used. 
The uniform matter symmetry energy obtained in this approximation goes 
linearly to zero when the density goes to zero,
  \begin{equation}
\label{linEsym}
E_{\rm sym}(n,T) \propto n.
 \end{equation}
Such a behavior is often seen in the results shown in the literature, but 
is incorrect because correlations are not included.

At low density the symmetry energy changes mainly because additional 
binding is gained in symmetric matter due to formation of clusters and 
pasta structures~\cite{Watanabe:2009vi}.  Therefore, the symmetry energy 
in the low-temperature limit has to be equal to the binding energy per 
nucleon associated with the strong interaction of the most bound nuclear 
cluster.  Theoretical calculations of the density dependence 
of the symmetry energy based on conventional mean-field approaches and 
ignoring cluster formation will fail to give the correct low-temperature, 
low-density limit to the symmetry energy.  The correct low-density limit 
can be recovered only if the formation of clusters is properly taken 
into account, as has previously been shown in Ref.~\cite{Ropke:1983},
see also~\cite{Horowitz:2005nd,Schmidt:1990} in the context of a virial 
expansion valid at very low densities, and in Ref.~\cite{Typel:2009sy}.

Approaches used to account for cluster formation include the nuclear 
statistical equilibrium model (NSE)~\cite{Bondorf:1995ua}, cluster-virial 
expansions~\cite{Horowitz:2005nd}, as well as generalized Beth-Uhlenbeck 
approaches~\cite{Schmidt:1990}.  A thermodynamic Green's function approach that 
allows a generalization of the NSE model by introducing a quasiparticle 
description also for the bound states was already formulated some decades 
ago~\cite{Ropke:1983}, but only recently analyzed with respect to the 
experimental consequences for nuclear matter~\cite{Ropke:2008qk}. 

To deal with the clusterization, we employ a quantum statistical (QS) 
approach which takes into account cluster correlations in the medium. 
To extend the range of applicability of this approach, we then 
interpolate between the exact low-density limit and the very successful 
RMF approaches near the saturation density to provide a representation 
useful over a wide range of densities.

In this QS approach cluster correlations are described in a generalized 
Beth-Uhlenbeck expansion.  The advantage of this method is that the medium 
modifications of the clusters at finite density are taken into account.  
In Ref.~\cite{Typel:2009sy} the thermodynamic properties of nuclear matter 
were derived using this approach.  The formulation of Ref.~\cite{Typel:2009sy}
is valid in the density and temperature range where the formation of light 
clusters with $A \leq 4$ dominates and heavier clusters are not yet important. 
 The method requires a reasonably accurate modeling of the quasiparticle 
properties. For that  we employ a RMF model with density dependent 
couplings~\cite{Typ2005} which gives a good description both of nuclear 
matter around normal density and of ground state properties of nuclei 
across the nuclear chart.  In order to extend the applicability of this 
RMF model to very low densities, it has been generalized in 
Ref.~\cite{Typel:2009sy} to account also for cluster formation and 
destruction. The model allows derivation of the composition and the 
thermodynamic quantities of nuclear matter can be modeled in a large 
region of densities, temperatures and asymmetries that are required, 
for example, in supernova simulations.

\begin{figure}
\vspace {-0.75truein}
\resizebox{0.75\textwidth}{!}{%
\hspace{-3.0truein}
  \includegraphics{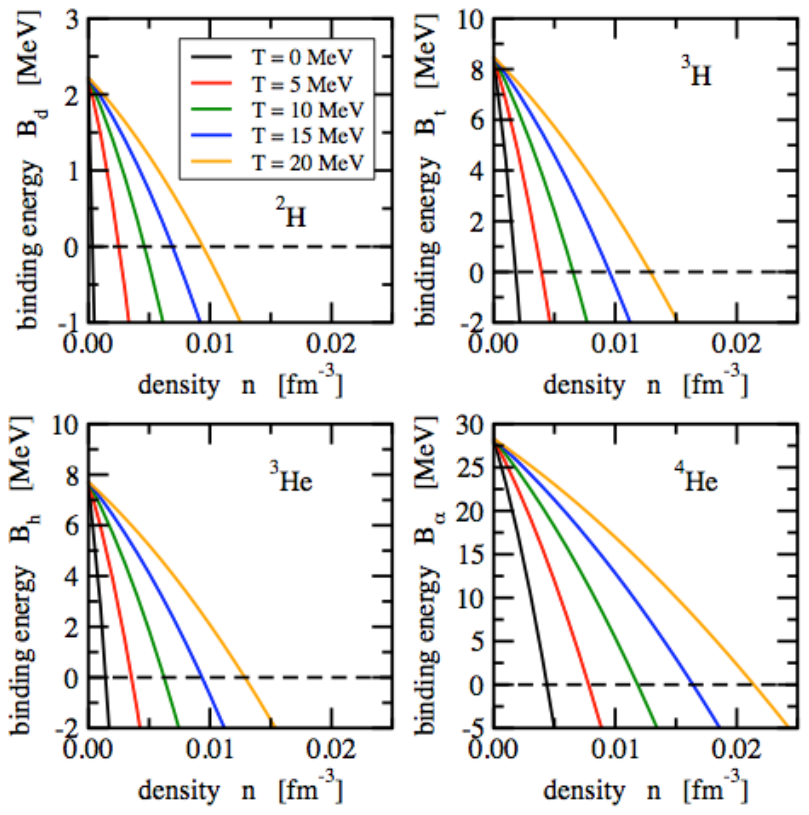}
}
\vspace {-0.75truein}
\caption{Change of the binding energy $B_i = -E^0_i$ of the clusters 
$i = d, t, h, \alpha$ at rest in symmetric nuclear matter due to the binding 
energy shift $B_i$ as used in the generalized RMF model as a function of 
the total nucleon density $n$. Mott points for specific temperature-density 
combinations appear when the binding energy relative to the medium 
becomes = 0.
}
\label{fig:2}       
\end{figure}

This generalized model naturally leads to a decrease of the cluster mass 
fractions at high densities, reflecting a reduction of the cluster binding 
energies due to Pauli blocking. The binding energy of a cluster relative to 
the medium vanishes at a point known as the Mott point. As a result, 
well-defined clusters appear  for densities below approximately 1/10 of 
the saturation density and get dissolved at higher densities. (Because 
of the presence of strong correlations in the scattering state continuum 
that are effectively represented by one resonance, there is a diminishing 
but non vanishing cluster fraction above the Mott density.) The Mott 
point is temperature dependent as seen in Figure~\ref{fig:2} where 
calculated Mott points for $d, t,  ^3$He and $\alpha$ particles are 
represented~\cite{Typel:2009sy}. 

\section{Quantum statistical approach to the symmetry energy of nuclear matter at 
finite temperature} 

\subsection{General approach}

To describe the thermodynamic properties of a many-particle system, the 
thermodynamic (Matsubara) Green's function approach can be used.  Exact 
expressions for the equations of state (\ref{density}) can be given that 
contain the single-particle spectral function $S_\tau({\bf p},\omega)$,

Using the finite-temperature Green function formalism, 
a non-relativistic quantum statistical approach 
can be given to describe the equation of state of nuclear matter
including the formation of bound states~\cite{Ropke:1983,Schmidt:1990}.
It is most convenient to start with the nucleon number densities 
$n_{\tau}(T,{\mu}_{p},{\mu}_{n})$ as
functions of temperature 
$T$ and non-relativistic 
chemical potentials ${\mu}_{\tau}$ 
for protons ($\tau = p$) and neutrons ($\tau = n$), respectively,
\begin{equation}
 n_{\tau}(T,{\mu}_p,{\mu}_n)=
 2  \int \frac{d^{3}k_{1}}{(2 \pi)^{3}}
 \int_{-\infty}^\infty \frac{d \omega}{2 \pi} f_{1,Z}(\omega) 
 S_{\tau}(1,\omega)\,. 
\label{eosspec}
\end{equation}
Summation over spin yields the factor $2$ and 
\begin{equation}
\label{faz}
 f_{A,Z}(\omega) = \left(\exp\left\{ 
\beta \left[\omega -Z {\mu}_{p} -(A-Z) {\mu}_{n}\right] 
 \right\} - (-1)^{A}\right)^{-1}
\end{equation}
is the Fermi or Bose distribution function which depends on the inverse 
temperature $\beta = 1/T$. Instead of the isospin quantum number $\tau$ 
we occasionally use the mass number $A$ and the charge number $Z$.
Both the distribution function and the spectral function $S_1(1,\omega)$ 
depend on the temperature and the chemical potentials ${\mu}_{p}$, 
${\mu}_{n}$ not given explicitly.  For this EOS (\ref{eosspec}), expressions 
such as the Beth-Uhlenbeck formula and its generalizations have been 
derived~\cite{Horowitz:2005nd,Ropke:1983,Schmidt:1990}.

The spectral function $S_{\tau}(1,\omega)$ 
is related to the self-energy $\Sigma(1,z)$ according to
\begin{equation}
\label{spectral}
 S_{\tau}(1,\omega) = \frac{2 {\rm Im}\,\Sigma(1,\omega-i0)} 
{[\omega - E(1)- {\rm Re}\, \Sigma(1,\omega)]^2 + 
[{\rm Im}\, \Sigma(1,\omega-i0)]^2 } \: ,
\end{equation}
where the imaginary part has to be taken for a small negative imaginary part 
in the frequency $\omega$. $E(1)=p^{2}/(2m)$ is the
kinetic energy of the free nucleon.
The solution of the relation 
\begin{equation}
\label{quasinucleon}
E_{1}^{\rm qu}(1) = E(1) + {\rm Re}\, \Sigma[1,E_{1}^{\rm qu}(1)] 
\end{equation}
defines the single-nucleon quasiparticle energies 
$E_{1}^{\rm qu}(1) = E(1) + \Delta E^{\rm SE}(1)$. 
Expanding for small Im~$\Sigma(1,z)$, the spectral function yields a 
$\delta$-like contribution. The densities
are calculated from Fermi
distributions with the quasiparticle energies so that
\begin{equation}
 n^{\rm qu}_{\tau}(T,{\mu}_{p},{\mu}_{n})
 =\frac{2}{V} \sum_{k_{1}} f_{1,Z}[E_{1}^{\rm qu}(1)] 
\label{nqu}
\end{equation}
follows for the EOS in mean field approximation.
This result does not contain the contribution of bound states and therefore 
is not correct in the low-temperature, low-density limit where the NSE 
describes the nuclear matter EOS.

As shown in Refs.~\cite{Ropke:1983,Schmidt:1990}, the bound state contributions
are obtained from the poles of Im~$\Sigma(1,z)$ which cannot be neglected in 
expanding the spectral function with respect to Im~$\Sigma(1,z)$.  A cluster 
decomposition of the self-energy has been proposed, see~\cite{Ropke:1983}.
The self-energy is expressed in terms of the $A$-particle Green functions which
read in bilinear expansion
\begin{eqnarray}
 &&G_A(1 \dots A,1^{\prime}\dots A^{\prime},z_A)
  =\sum_{\nu K} \psi_{A \nu K}(1\dots A)\nonumber\\ 
  && \times
 \frac{1}{ z_{A}-E^{\rm qu}_{A,\nu}(K)} 
 \psi^{\ast}_{A \nu K}(1^{\prime}\dots A^{\prime}) \: .
\label{bilinear}
\end{eqnarray}
The $A$-particle wave function $\psi_{A \nu K}(1\dots A) $
and the corresponding eigenvalues $E^{\rm qu}_{A, \nu}(K)$ result from solving 
the in-medium Schr\"{o}dinger equation, see the following Subsections.
$K$ denotes the center of mass momentum of the $A$-nucleon system. 
Besides the bound states, the summation over the internal quantum states $\nu$ 
includes also the scattering states.

The evaluation of the equation of state in the low-density limit is
straight forward.
Considering only the bound-state contributions, we obtain the result 
\begin{eqnarray}
\label{EoS:p}
 n^{\rm tot}_{p}(T,{\mu}_p,{\mu}_n)&=& 
 \frac{1 }{ V} \sum_{A,\nu,K}Z 
 f_{A,Z}[E^{\rm qu}_{A,\nu}(K;T,{\mu}_p,{\mu}_n)]\,,
 \\  
 n^{\rm tot}_{n}(T,{\mu}_p,e{\mu}_n)&=& 
 \frac{1}{ V} \sum_{A,\nu,K}(A-Z) 
 f_{A,Z}[E^{\rm qu}_{A,\nu}(K;T,{\mu}_p,{\mu}_n)]\,\nonumber 
\label{quasigas}
\end{eqnarray}
for the EOS describing a mixture of components (cluster quasiparticles) obeying
Fermi or Bose statistics. 
The total baryon density results as 
$n(T,{\mu}_{p},{\mu}_{n})
= n^{\rm tot}_{n}(T,{\mu}_{p},{\mu}_{n})
 +n^{\rm tot}_{p}(T,{\mu}_{p},{\mu}_{n})$. 
To derive the extended Beth-Uhlenbeck formula, see~\cite{Schmidt:1990},
we restrict the
summation to $A \le 2$, but extend the summation over the internal quantum
numbers $\nu$,
not only to the excited states, but also the scattering states.
Note that at low temperatures Bose-Einstein condensation may occur.

The NSE is obtained in the low-density limit if the in-medium energies  
$E^{\rm qu}_{A,\nu}(K;T,{\mu}_{p},{\mu}_{n})$ 
can be replaced by the binding energies of the isolated nuclei 
$E^{(0)}_{A,\nu}(K)=E_{A,\nu}^{(0)}+K^{2}/(2Am)$, with
$m=939$~MeV the average nucleon mass. 
For the cluster contributions, i.e.\  $A>1$, the summation
over the internal quantum numbers is again restricted to the bound states only.
We have
\begin{eqnarray}
\label{NSE:p}
 n^{\rm NSE}_{p}(T,{\mu}_p,{\mu}_n)&=& 
 \frac{1 }{V} \sum_{A,\nu,K}^{\rm bound} Z 
 f_{A,Z}[E^{(0)}_{A,\nu}(K)]\,,
  \\  
 n^{\rm NSE}_{n}(T,{\mu}_p,{\mu}_n)&=& 
 \frac{1}{V} \sum_{A,\nu,K}^{\rm bound} (A-Z) 
 f_{A,Z}[E^{(0)}_{A,\nu}(K)]\,.\nonumber
\end{eqnarray}
The summation over $A$ includes also the contribution of
free nucleons, $A=1$.

In the nondegenerate and nonrelativistic case assuming a Maxwell-Boltzmann 
distribution, the summation over the momenta $K$ can be performed 
analytically and the thermal wavelength $\Lambda= \sqrt{2 \pi/(m T)}$ of 
the nucleon occurs.  As shown below, the medium effects in nuclear matter 
are negligible below 10$^{-4}$ times the saturation density $n_0$ 
for the temperatures considered here.

Interesting quantities are the mass fractions
\begin{equation}
X_{A,Z}= \frac{A}{V n} \sum_{\nu, K}
 f_{A,Z}[E^{\rm qu}_{A,\nu}(K;T,{\mu}_p,{\mu}_n)]
\end{equation}
of the different clusters.  From the EOS considered here, thermodynamical 
potentials can be obtained by integration, in particular the free energy 
per volume $f={\tilde F}/V$. In the special case of symmetric nuclear matter,
$Y_p^{\rm s}=0.5$, the free energy per volume is obtained from the averaged 
chemical potential $ \mu =( \mu_p+ \mu_n)/2$ (symmetric matter) as
\begin{equation}
\label{freV}
f(T,n,Y_p^{\rm s})  = \int_0^n dn' \:
  \mu(T,n',Y_p^{\rm s}) \:  .
\end{equation}

In the quantum statistical approach described above,
we relate the EOS to properties of the correlation functions, in particular to
the peaks occurring in the $A$-nucleon spectral function describing the 
single-nucleon quasiparticle ($A=1$) as well as the nuclear quasiparticles 
($A \ge 2$).  
The microscopic approach to these quasiparticle energies can be given 
calculating the self-energy. 
Different approaches can be designed which reproduce known properties of the 
nucleonic system. 

\subsection{\label{subsec:medium1}Medium modification 
of single nucleon properties}

The single-particle spectral function contains the
single-nucleon quasiparticle contribution,
$E_{1}^{\rm qu}(1) = E^{\rm qu}_{\tau}(k)$, given in Eq.\ 
(\ref{quasinucleon}), 
where $\tau$ denotes isospin of particle $1$ and $ k$ is the momentum.
In the effective mass approximation, the single-nucleon quasiparticle
dispersion relation reads
\begin{equation}
\label{quasinucleonshift}
E_{\tau}^{\rm qu}(k) = \Delta E^{\rm SE}_{\tau}(0) +\frac{k^2}{2
m_\tau^{\ast}} + {\mathcal O}(k^4)\,,
\end{equation}
where the quasiparticle energies are shifted at zero momentum $ k$
by $\Delta E^{\rm SE}_{\tau}(0)$, and $m_\tau^{\ast}$ 
denotes the effective mass of neutrons
($\tau=n$) or protons ($\tau=p$).
Both quantities, $\Delta E^{\rm SE}_{\tau}(0)$ and $m_\tau^{\ast}$, are
functions of $T$, $n_{p}$ and $n_{n}$, characterizing the surrounding matter.

Expressions for the  single-nucleon quasiparticle energy 
$E^{\rm qu}_{\tau}(k)$ can be given by the Skyrme 
parametrization~\cite{Vautherin:1971aw} or by more sophisticated approaches 
such as relativistic mean-field (RMF) approaches~\cite{Typ2005}, and 
relativistic Dirac-Brueckner Hartree-Fock~\cite{Fuchs:2005yn} calculations.  
We use the density-dependent relativistic mean field approach 
of~\cite{Typ2005} that is designed not only to reproduce known properties 
of nuclei, but also fits with microscopic calculations in the low density
region. 

We can assume~\cite{Klahn:2006ir} that 
the density-dependent RMF parametrisation covers a large density 
region. It will be used in this work to determine the single-nucleon 
quasiparticle energies. 
The  single-nucleon quasiparticle energies result as
\begin{equation}
E^{\rm qu}_{n,p}(0)=\sqrt{[m^2 - 
                 S(T,n,\pm\delta)]^2 + k^2} + V(T,n,\pm\delta) \: .
\end{equation}
In the nonrelativistic limit, the shifts of the quasiparticle energies are
\begin{equation}
\Delta E^{\rm
     SE}_{n,p}(k)=V(T,n,\pm\delta)-S(T,n,\pm\delta)\: .
\end{equation}
The effective masses for neutrons and protons are given by
\begin{equation}
m_{n,p}^{\ast} = m - S(T,n,\pm\delta) \: .
\end{equation}
Approximations for the functions $V(T,n,\delta)$ and $S(T,n,\delta)$ are 
given in the literature~\cite{Roe13}.  These functions reproduce the 
empirical values for the saturation density 
$n_0\approx 0.15$~fm$^{-3}$ and the binding energy per nucleon
$B/A\approx -16$~MeV.  The effective mass is somewhat smaller than the 
empirical value $m^{\ast} \approx m(1-0.17~n/n_0)$ 
for $n<0.2$ fm$^{-3}$.

\subsection{\label{subsec:medium2}Medium modification of cluster properties}

Recent progress of the description of clusters in low density nuclear
matter~\cite{Sumiyoshi:2008,Ropke:2008qk,Ropke:2005}
enables us to evaluate the properties 
of deuterons, tritons, helions
and helium nuclei in a non-relativistic 
microscopic approach, taking the influence of the
medium into account. 

In addition to the $\delta$-like nucleon quasiparticle contribution, also the
contribution of the bound and scattering states
can be included in the single-nucleon spectral function by analyzing the
imaginary part of $\Sigma (1,z)$. Within a cluster decomposition,
$A$-nucleon $T$ matrices appear in a many-particle approach. These $T$
matrices describe the propagation of the $A$-nucleon cluster in nuclear
matter. In this way, bound states contribute to  $n_{\tau} =
n_{\tau}(T,\mu_{n},\mu_{p})$, 
see~\cite{Ropke:1983,Schmidt:1990}. 
Restricting the cluster decomposition only to the contribution of two-particle 
correlations, we obtain the so-called $T_{2}G$ approximation. 
In this approximation, the Beth-Uhlenbeck formula is obtained for the EOS, as 
shown in~\cite{Ropke:1983,Schmidt:1990}. 
In the low-density limit, the propagation of the $A$-nucleon cluster is 
determined by the energy eigenvalues of the corresponding nucleus, and the 
simple EOS (\ref{EoS:p}) results describing the nuclear statistical 
equilibrium (NSE).

For nuclei imbedded in nuclear matter, an effective wave equation
can be derived~\cite{Ropke:1983,Ropke:2008qk}.
The $A$-particle wave function $\psi_{A\nu K}(1\dots A)$
and the corresponding eigenvalues $E^{\rm qu}_{A, \nu}(K)$
follow from solving the in-medium Schr\"{o}dinger equation 
\begin{eqnarray}
\lefteqn{[E^{\rm qu}(1)+\dots + E^{\rm qu}(A) - E^{\rm qu}_{A, \nu}(K)]
 \psi_{A\nu K}(1\dots A)}
\nonumber \\ &&
+\sum_{1^{\prime} \dots A^{\prime}}\sum_{i<j}[1-\tilde{f}(i)- \tilde{f}(j)]
 V(ij,i^{\prime}j^{\prime}) \nonumber \\ && \times \prod_{k \neq
  i,j} \delta_{kk^{\prime}}\psi_{A \nu K}(1^{\prime}\dots A^{\prime})=0\,.
\label{waveA}
\end{eqnarray}
This equation contains the effects of the medium in the single-nucleon
quasiparticle shifts as well as in the Pauli blocking terms. The 
$A$-particle wave function and energy depend on the total momentum
$K$ relative to the medium.

The in-medium Fermi distribution function 
$\tilde{f}(1)=\left(\exp\left\{ \beta 
\left[E^{\rm qu}(1)- {\tilde \mu}_{\tau_1}\right]\right\} +1 \right)^{-1}$ 
contains the non-relativistic
effective chemical potential ${\tilde \mu}_{\tau}$ which is determined
by the total proton or neutron density calculated in quasiparticle
approximation,
$n_{\tau}^{\rm tot} = 
V^{-1} \sum_{1} \tilde f(1) \delta_{\tau_{1},\tau}$ for 
the particles inside the volume $V$. It describes the
occupation of the phase space neglecting any correlations in the medium.
The solution of the in-medium Schr\"odinger equation (\ref{waveA}) can
be obtained in the low-density region by perturbation theory. In
particular, the quasiparticle energy of the $A$-nucleon cluster with
$Z$ protons in the ground state follows as
\begin{eqnarray}
&&E^{\rm qu}_{A,\nu}(K) = 
 E_{A,Z}^{(0)}+\frac{ K^{2}}{2 A m} \nonumber \\ &&+
\Delta E_{A,Z}^{\rm SE}(K)+\Delta E_{A,Z}^{\rm Pauli}(K) 
 + \Delta E_{A,Z}^{\rm Coul}(K), 
\label{finalshift}
\end{eqnarray}
plus higher order contributions with respect to density. Besides the cluster binding energy in the vacuum
$E_{A,Z}^{(0)}$
and the kinetic term, the self-energy shift 
$\Delta E_{A,Z}^{\rm SE}(K)$, the Pauli shift $\Delta E_{A,Z}^{\rm
  Pauli}(K)$
and the Coulomb shift $\Delta E_{A,Z}^{\rm
Coul}(K)$ enter. The latter
can be evaluated for dense matter in the Wigner-Seitz
approximation~\cite{Ropke:1984,Kolomiets:1997zzb,Shlomo:2005}.
It is given by 
\begin{equation}
\Delta E_{A,Z}^{\rm Coul}(K)=\frac{Z^2}{A^{1/3}} 
\frac{3}{5} \frac{e^{2}}{r_{0}} 
\left[ \frac{3}{2} \left(\frac{2 n_p}{n_0}\right)^{\frac{1}{3}} 
- \frac{n_p}{n_0}\right]  
\end{equation}
with $r_{0} = 1.2$~fm. Since the values of $Z$ are small, this 
contribution is small as well and disregarded here together with other 
small higher order terms in the quasiparticle energy (\ref{finalshift}).

The self-energy contribution to the quasiparticle shift is determined by the
contribution of the single-nucleon shift
\begin{equation}
\label{delArigid}
\Delta E_{A,Z}^{\rm SE}(0)= (A-Z) \Delta E_{n}^{\rm SE}(0)+ Z \Delta
E_{p}^{\rm SE}(0) +\Delta E_{A,Z}^{\rm SE, eff.mass}\: .
\end{equation}
The contribution to the self-energy shift due to the change of the effective 
nucleon mass can be calculated from perturbation theory
using the unperturbed wave function of the clusters, 
see~\cite{Sumiyoshi:2008}, so that
\begin{equation}
\label{delSEeff}
\Delta E_{A,Z}^{\rm SE, eff.mass}=\left(\frac{m}{m^{\ast}}-1\right)s_{A,Z}\: .
\end{equation}
Values of $s_{A,Z}$
for $ \{A,Z\}=\{i\}=\{d,t,h,\alpha\} $ are given in Ref.~\cite{Typel:2009sy}.
Inserting the medium-dependent quasiparticle energies in the
distribution functions~(\ref{faz})
this contribution to the quasiparticle shift can
be included renormalizing the chemical potentials.

The most important effect  in
the calculation of the abundances of light
elements comes from the Pauli blocking terms in Eq.\ (\ref{waveA}) in
connection with the interaction potential. This contribution is
restricted only to the bound states so that it may lead to the
dissolution of the nuclei if the density of nuclear matter increases.
The corresponding shift $\Delta E_{A,Z}^{\rm Pauli}(K)$ can be evaluated
in perturbation theory provided the interaction potential and the
ground state wave function are known.
After angular averaging where in the Fermi functions the mixed 
scalar product $\vec{k} \cdot \vec{K}$ between the total momentum 
 $\vec{K}$ 
and the remaining Jacobian coordinates $\vec{k}$ is neglected, 
the Pauli blocking shift can be approximated as
\begin{equation}
\label{delpauli0P1}
\Delta E_{A,Z}^{\rm Pauli}(K) \approx \Delta E_{A,Z}^{\rm Pauli}(0) \,
\exp\left(-\frac{ K^{2}}{g_{A,Z}}\right) 
\end{equation}
where the amplitude $\Delta E_{A,Z}^{\rm Pauli}(0)$ and the dispersion $ g_{i}$
depend on the thermodynamic parameters $(T,n,Y_{p})$.
Values are given in~\cite{Typel:2009sy}, for a more recent calculation 
see~\cite{Ropke:2008qk,Ropke2011}.

With the neutron number $N_{i} = A_{i}-Z_{i}$, 
it can be written as
\begin{equation} \label{eq:lin_be_shift}
\Delta E_{A_{i},Z_{i}}^{\rm
  Pauli}(0;n_{p},n_{n},T) = 
 -\frac{2}{A_{i}}
 \left[Z_{i}n_{p}+N_{i}n_{n}\right] 
 \delta  E^{\rm
  Pauli}_{i}(T,n)~,
\end{equation}
where the temperature dependence and 
higher density corrections are contained in the functions
$ \delta  E^{\rm
  Pauli}_{i}(T,n)$. 

Now, the nucleon number densities 
(\ref{quasigas}) can be evaluated as in the non-interacting case, 
with the only difference that
the number densities of the particles are calculated with the 
quasiparticle energies.
In the light cluster-quasiparticle approximation, the total densities of 
neutrons (note that we change the notation to distinguish between the total 
nucleon density, former $n_\tau$, and the free nucleon density)
\begin{equation}
 n_{n}^{\rm tot} = n_{n} + \sum_{i=d,t,h,\alpha} N_{i} n_{i}
\end{equation}
and of protons
\begin{equation}
 n_{p}^{\rm tot} = n_{p} + \sum_{i=d,t,h,\alpha} Z_{i} n_{i}
\end{equation}
contain the densities of the free neutrons and protons 
$n_{n}$ and $n_{p}$,
respectively, and the contributions from the nucleons bound in the
clusters with densities $n_{i}$. 
The state of the system in chemical equilibrium
is completely determined by specifying the
total nucleon density $n=n_{n}^{\rm tot}+n_{p}^{\rm tot}$,
the asymmetry $\delta$ 
and the temperature $T$ as long as no $\beta$-equilibrium is considered.

This result is an improvement of the NSE and allows for the smooth
transition from the low-density limit up to the region of saturation
density. The bound state contributions to the EOS are fading with
increasing density because they move as resonances into 
the continuum of scattering states. This improved NSE, however, does not 
contain the contribution of scattering states explicitly.  For the treatment 
of continuum states in the two-nucleon case, as well as the evaluation of 
the second virial coefficient, see~\cite{Horowitz:2005nd,Schmidt:1990}.

The account of scattering states needs further consideration. 
Investigations on the two-particle level have been performed and extensively 
discussed~\cite{Horowitz:2005nd,Ropke:1983,Schmidt:1990}.
We use the Levinson theorem to take the contribution of scattering 
states into account in the lowest-order
approximation. Each bound state
contribution to the density has to accompanied with a continuum
contribution that partly compensates the strength of the bound state
correlations.
As a consequence, the total proton and neutron densities are given by
\begin{eqnarray}
\label{quasigas2_p}
 n^{\rm tot}_{p}(T,{\mu}_p,{\mu}_n)&=& 
 \frac{1 }{ V} \sum_{A,\nu,K}^{\rm bound} Z 
\left[  f_{A,Z}[E^{\rm qu}_{A,\nu}(K;T,{\mu}_p,{\mu}_n)] \right. \nonumber \\&& \left.
- f_{A,Z}[E^{\rm cont}_{A,\nu}(K;T,{\mu}_p,{\mu}_n)] \right]\,,
  \\  
\label{quasigas2_n}
 n^{\rm tot}_{n}(T,\tilde{\mu}_p,\tilde{\mu}_n)&=& 
 \frac{1}{V} \sum_{A,\nu,K}^{\rm bound} (A-Z) 
\left[  f_{A,Z}[E^{\rm qu}_{A,\nu}(K;T,{\mu}_p{\mu}_n)] \right. \nonumber \\&& \left.
- f_{A,Z}[E^{\rm cont}_{A,\nu}(K;T{\mu}_p,{\mu}_n)] \right]\,
\end{eqnarray}
with explicit bound and scattering terms.
 $E^{\rm cont}_{A,\nu}$ denotes the edge of the 
continuum states that is also determined by the single-nucleon 
self-energy shifts. 
These expressions guarantee a smooth behavior if bound 
states merge with the continuum of scattering states.
The summation over $A$ includes also the contribution of free nucleons, $A=1$, 
considered as quasiparticles with the energy dispersion 
given by the RMF approach. 

The summation over $K$ and the subtraction of the
continuum contribution is extended only over 
that region of momentum space where 
bound states exist. The disappearance 
of the bound states is caused by the Pauli blocking term; the self-energy
contributions to the quasiparticle shifts act on bound as well 
as on scattering states.
Above the so-called Mott density, where the bound states at $K = 0$ disappear,
the momentum summation has to be extended only over that region
$K > K^{\rm Mott}_{A,\nu}(T,n,\delta)$ where the bound state energy 
is lower than the continuum of scattering states.  The contribution of 
scattering states is necessary to obtain the second virial coefficient 
according to the Beth-Uhlenbeck equation, 
see~\cite{Horowitz:2005nd,Schmidt:1990}.  This leads also to corrections in 
comparison with the NSE that accounts only for the bound state contributions,
neglecting all effects of scattering states. These corrections become 
important at increasing temperatures for weakly bound clusters. Thus,
the corrections which lead to the correct second virial coefficient are
of importance for the deuteron system, when the temperature is
comparable or large compared with the binding energy per nucleon. In the
calculations for the quantum statistical (QS) model
shown below, the contributions of these continuum
correlations have been taken into account.

Solving Eqs. (\ref{quasigas2_p}) and (\ref{quasigas2_n})
for given $T$, $n^{\rm tot}_{p}$ and 
$n^{\rm tot}_{n}$ we find the
chemical potentials $\mu_p$ and $\mu_n$. After integration, 
see Eq. (\ref{freV}), the free
energy is obtained, and all the other thermodynamic 
functions are derived from this quantity
without any contradictions. Results are given below.

We do not consider the formation of heavy clusters here.  This limits the 
parameter range $n_{n}^{\rm tot}$, $n_{p}^{\rm  tot}$, $T$ in the phase 
diagram to that area where the abundances of heavier clusters are small. 
For a more general approach to the EOS which takes also the contribution of 
heavier cluster into account, see~\cite{Ropke:1984}.  Future work will 
include the contribution of the heavier clusters. 

Further approximations refer to the linear dependence on density of the shifts 
of binding energies, calculated in perturbation theory.  A better treatment 
will improve these shifts, but it can be shown that the changes are small.  
Furthermore, the approximation of the uncorrelated medium can be improved 
considering the cluster mean-field 
approximation~\cite{Ropke:1983,Ropke:2008qk,Ropke:1980}.  Last but not 
least, the formation of quantum condensates will give further contributions 
to the EOS. However, in the region considered here the formation of quantum 
condensates does not appear. 

\subsection{Results for the symmetry energy}

\begin{figure}
\resizebox{0.545\textwidth}{!}{%
  \includegraphics{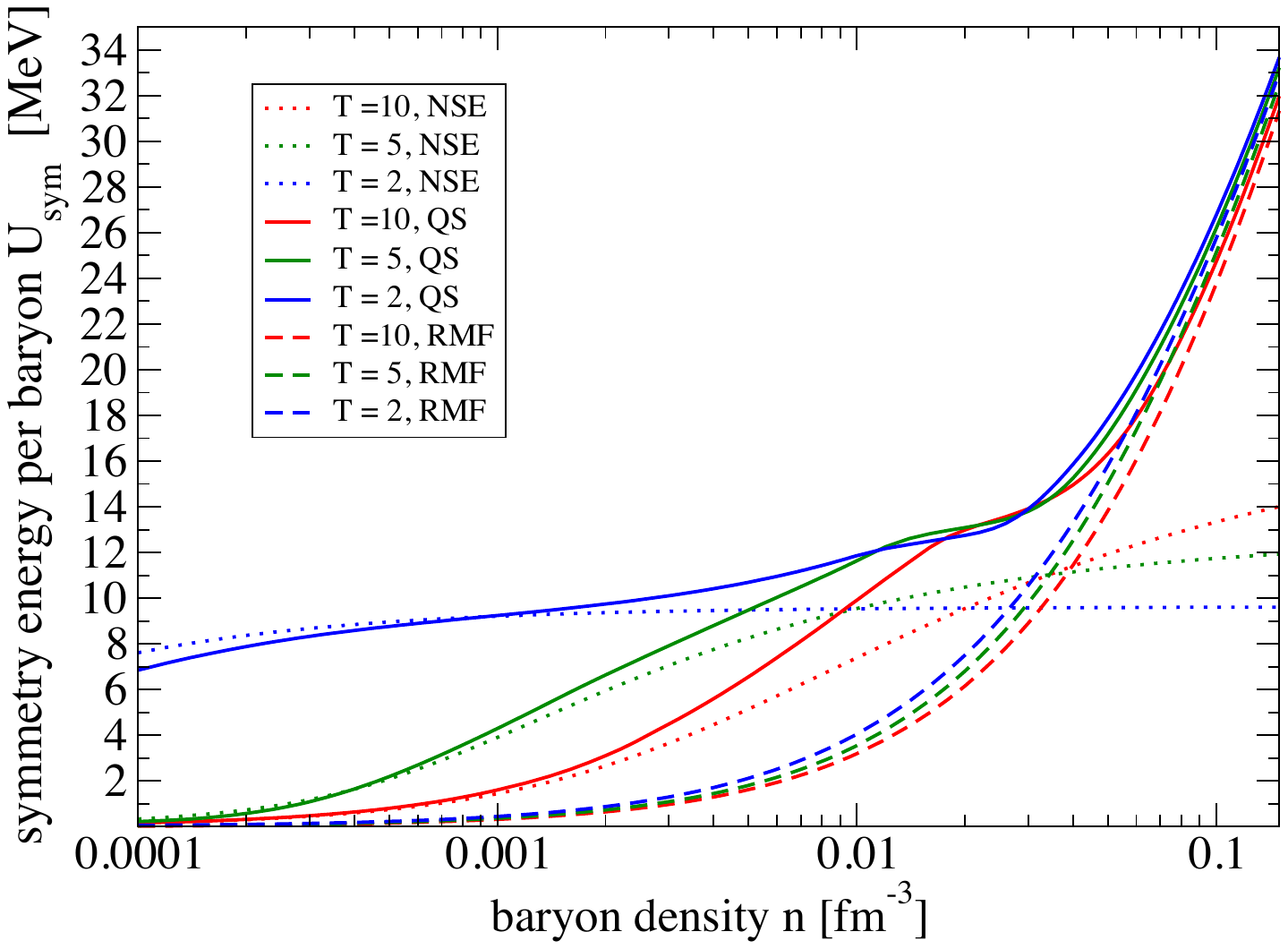}
}
\caption{QS calculation of internal symmetry energy as function of baryon 
density $n$ for different temperatures.  For comparison, the NSE and RMF 
approximations are also given.}
\label{fig:3}       
\end{figure}

\begin{figure}
\resizebox{0.545\textwidth}{!}{%
  \includegraphics{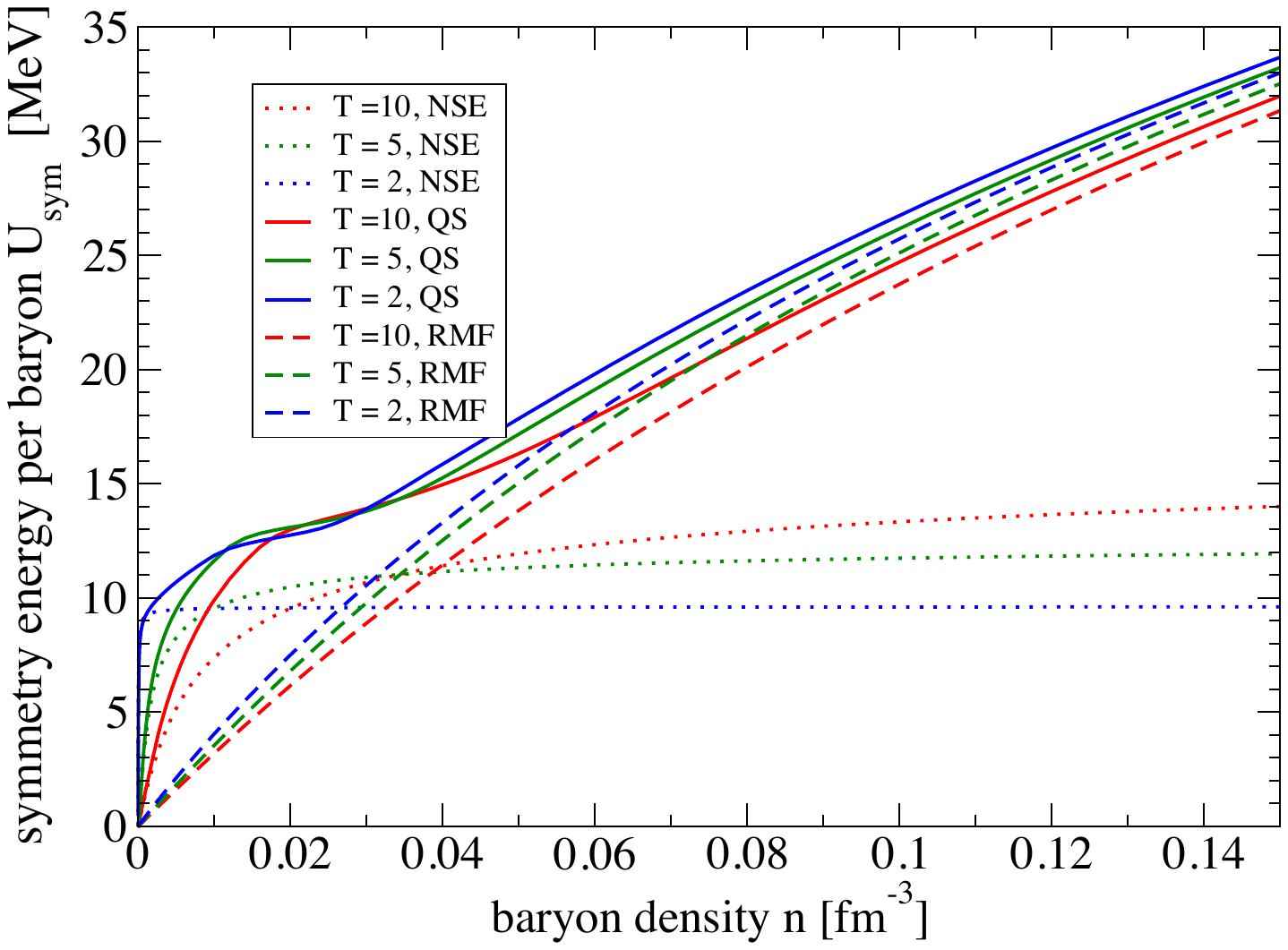}
}
\caption{The same as Fig.~\ref{fig:3}, linear density scale.}
\label{fig:4}       
\end{figure}

We perform the calculations for $A \le 4$ and compare them with several
approximations. Solving the EOS~(\ref{quasigas2_p}), (\ref{quasigas2_n}), we 
find the chemical potentials for symmetric matter as function of the 
densities and the temperature. The free energy density is obtained after 
integration, and the internal energy follows from the standard thermodynamic 
relations.  The same procedure is made for neutron matter, and the difference 
of the internal energies gives the symmetry energy as shown in 
Figures~\ref{fig:3} and~\ref{fig:4}.  The symmetry energy is calculated by 
using the finite difference formula, Eq. (12).  The Coulomb energy of the 
clusters has been subtracted.  Cluster formation is more clearly seen in the 
logarithmic scale, Fig.~\ref{fig:3}, whereas the linear scale, 
Fig.~\ref{fig:4}, is commonly used.  The disappearance of clusters 
around $n_0/5$ indicates the transition to the RMF result.  Because of the 
temperature dependence of the composition, the symmetry energy that is 
determined by the yields of clusters also depends strongly on $T$. 
 
Degeneration effects are small, but can be easily incorporated.  At low 
temperatures and high densities nucleons follow Fermi statistics. 
Pairing occurs in the two-nucleon channel and the transition from Cooper 
pairing to Bose-Einstein condensation is included.  Even more interesting 
is Bose condensation for larger bosonic bound states that may occur at 
very low temperatures.  The quarteting is a possible contribution that 
is presently not included.

This analysis should be improved by further considering the  effects of
continuum correlations. In particular, this will affect the contribution 
of deuterons that are weakly bound, and has been already taken into account 
in the present calculation.  The Beth-Uhlenbeck formula gives the 
exact expression for the second virial coefficient.  The effects become 
more relevant for high temperatures.

Due to the formation of correlations, in particular clusters, the symmetry 
energy becomes strongly temperature dependent. For decreasing temperatures, 
the contribution of higher clusters $A > 4$ is increasing.  For an estimate 
of the contribution of higher clusters see~\cite{Typel2013,Hempel:2010}. 
Charge-neutral nuclear matter will not show a first order phase transition 
because of the Coulomb interaction that gives diverging energy, if with a 
homogeneous  background of electrons the nuclear matter disensembles into 
a liquid phase and a gas-like phase. In that region, droplets and 
pasta-like structures may be formed, that are not fixed in space. Thus, 
the system remains in a homogeneous state in the thermodynamic context with 
a constant average density.
               
\section{Laboratory Tests of the Equation of State at Low Density}
\subsection{Experimental Technique}
 
The experimental information is derived from heavy-ion collisions of charge 
asymmetric nuclei, where transient states of different density can be 
reached, depending on the incident energy and the centrality of the 
collision.  In the Fermi energy domain effects of the symmetry energy have 
been investigated in judiciously chosen 
observables~\cite{Li:2008gp,Baran:2004ih} suggesting an almost linear behavior 
around and below normal density~\cite{Tsang:2008fd,Wuenschel:2009,Sfienti09}.

Our experimental investigations of low density nuclear matter have 
utilized near Fermi energy heavy ion collisions to produce heated and 
expanded matter~\cite{Kowalski:2006ju,wang07,Qin,Wada,natowitz10}.  
Cluster production was studied using the 4$\pi$ 
multi-detector, NIMROD, at the Cyclotron Institute at Texas A\&M University. 
NIMROD consists of a 166 segment charged particle array set inside a neutron 
ball~\cite{wuenschel09}.  The charged particle array is arranged in 12 rings 
of Si-CsI telescopes or single CsI detectors concentric around the beam 
axis. The CsI detectors are 1-10 cm thick Tl doped crystals read by 
photomultiplier tubes. A pulse shape discrimination method is employed to 
identify light particles in the CsI detectors. For this experiment each of 
the forward rings included two segments having two Si detectors 
(150 and 500 $\mu$m thick) in front of the CsI detectors 
(super telescopes) and three having one Si detector (300 $\mu$m thick). 
Each super telescope was further divided into two sections. Neutron 
multiplicity was measured with the 4$\pi$ neutron detector surrounding 
the charged particle array. This detector is a neutron calorimeter 
filled with Gd doped pseudocumene. Thermalization and 
capture of emitted neutrons in the ball leads to scintillation which 
is observed with phototubes providing event by event determination 
of neutron multiplicities.  Further details on the detection system, 
energy calibrations and neutron ball efficiency may be found in 
Ref.~\cite{hagel00,qin08}.  The combined neutron and charged particle 
multiplicities were employed to select the most violent events for 
subsequent analysis. 

Analyzing heavy ion collisions using the NIMROD multi-detector at 
Texas A\&M University, the medium modification of light fragments that 
leads to the dissolution has been shown.  Yields of light particles 
produced in the collisions of 47 $A$ MeV $^{ 40}$Ar with $^{112}$Sn ,
$^{124}$Sn and  $^{64}$Zn with $^{112}$Sn, $^{124}$Sn were employed in Thermal
coalescence model analyses to derive densities and temperatures
of the evolving emitting systems. Free symmetry energies of these systems 
were determined using Isoscaling analyses. 

The light particles and clusters emitted at the early stage of such 
collision are messengers carrying information on the dynamic evolution 
of the system and its degree of equilibration. We measure their energies, 
angles and yields and use that information to probe the properties of the 
system. We do this as a function of surface velocity of the emitted 
species.  The surface velocity $v_{\rm surf}$, i.e.\ the velocity before 
the final Coulomb acceleration, serves as a surrogate to follow the 
time evolution of the system~\cite{wang05}.

Experimental determinations of the symmetry energy at very low densities 
produced in heavy ion collisions of $^{64}$Zn on $^{92}$Mo and $^{197}$Au 
at $35$ MeV per nucleon have also been reported~\cite{Kowalski:2006ju}.
The surface velocity was also used as a measure of the time when the particles 
leave the source under different conditions of density and temperature. 
Only values of $v_{\rm surf} < 4.5$~cm/ns were included in that work since 
it was argued that the system does not reach equilibrium for 
higher $v_{\rm surf}$, see Tab.~I of Ref.~\cite{Kowalski:2006ju}.

Another experiment~\cite{10,40} was performed to study the dependence of the 
thermodynamic properties of nuclear material on neutron-proton 
asymmetry~\cite{AsyDepMcI}.  Here, collisions of $^{70}$Zn+$^{70}$Zn, 
$^{64}$Zn+$^{64}$Zn, and $^{64}$Ni+$^{64}$Ni at $E/A = 35$ MeV were 
studied. Charged particles and free neutrons produced in the 
reactions were measured in the NIMROD-ISiS 4$\pi$ detector 
array~\cite{wuenschel09,25}.  The energy resolution achieved 
allowed excellent isotopic resolution of 
charged particles up to $Z=17$.  For events in which all charged particles 
are isotopically identified, the QP (quasi-projectile, the primary excited 
fragment that exists momentarily after a non-central collision) 
was reconstructed using the charged particles and free neutrons. 
Thus, the reconstruction includes determination of the QP composition, 
both $A$ and $Z$.  In order to select thermally equilibrated QPs, the QPs 
are required to be on-average spherical, in a narrow range of shape 
deformation. The excitation energy of the QPs is determined from the 
transverse
kinetic energy of the charged particles, the Q-value of the 
QP breakup, the neutron multiplicity, and average neutron kinetic energy. 
This method of reconstruction has previously been described in 
detail~\cite{Wuenschel:2009,25,20,21,24,43,44}.  In this way, a set of 
quasi-projectiles was obtained, tightly selected on 
mass ($48 \le A_{QP} \le 52$), and which have properties consistent with 
thermal equilibration (i.e. selected on minimal shape deformation), and 
with known neutron-proton asymmetry.

\subsection{Thermodynamic parameter determination}
 
The characterization of properties of this low density matter necessarily 
begins with the determination of the temperature and density regions 
actually sampled in the collision.  Temperature determinations are 
relatively well in-hand as there is a long history of experiments 
focusing on temperature determination which we will not repeat here. See for 
example~\cite{hagel00,21,24,nebbia86,xu86,fabris87,hagel88,wada89,gonin89,chbihi91,nayak92,schwarz93,natowitz95,tsang96,majka97,huang97,xi98,xi99,natowitz02,souza09}.  In the work described in the following we have 
employed the Albergo method~\cite{Albergo85} which uses double 
isotope ratios to determine temperatures based on chemical equilibrium 
assumptions.  Thus, temperatures are determined using a H-He thermometer 
based on the double yield ratio of deuterons, tritons, $^3$He and $^4$He. 
Another method to extract temperatures is the analysis of fluctuations 
in the transverse momentum~\cite{Zheng}.
  
Accurate density determinations are inherently more difficult.  Among the 
different experimental approaches which have been explored to extract 
densities for systems below normal density are:
\begin{itemize}
\item the use of the Albergo NSE based 
relations~\cite{Kowalski:2006ju,Albergo85},
\item the use of  the Mekjian coalescence model which takes into 
account three body terms which might mimic either a higher density 
(three body collisions) or Pauli blocking~\cite{Qin,Mekjian},
\item analyses of caloric curve data or barrier data within the 
Fermi-Gas Model framework~\cite{Natowitz2,viola04},
\item the quantum fluctuation analysis method~\cite{Zheng},
\item an approach based on use of the  Chemical equilibrium  
constant employed in Ref.~\cite{Qin}.
\end{itemize}

\begin{figure}
\resizebox{0.545\textwidth}{!}{%
  \includegraphics{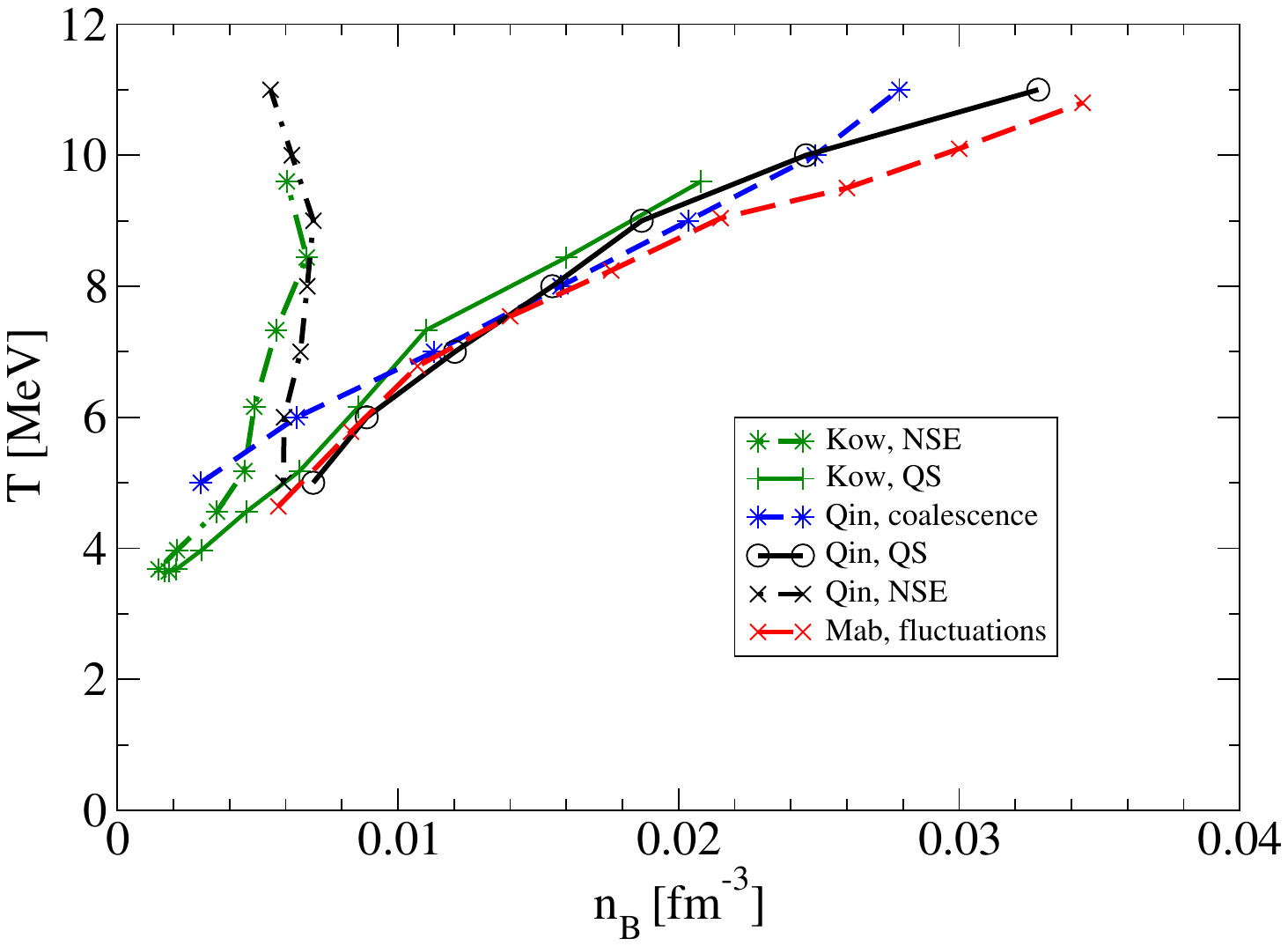}
}
\caption{Baryon density derived from yields of light elements. Data 
according to~\cite{Kowalski:2006ju} (Kow)~\cite{Qin}, (Qin)~\cite{Zheng},
(Mab) are compared with the results of the analysis of yields using NSE 
and QS calculations for $ K_c(\alpha)  $.}
\label{fig:5}       
\end{figure}

The last approach was described in a recent paper~\cite{R2013}.  The yields 
of $d, t, ^3$He and $^4$He for evolving intermediate source systems formed 
in the collisions of 47$A$ MeV  $^{40}$Ar with  $^{112}$Sn and $^{124}$Sn,  
and 47$A$ MeV     $^{64}$Zn with  $^{112}$Sn and $^{124}$Sn were determined  
and this technique  was applied to determine densities and temperatures. 
The free neutron yields are not measured but are determined from the free 
proton yield and the yield ratio of ${}^{3}$H to $^3$He. To determine
the asymmetry parameter of the emitting sources the total
proton and neutron yields including those bound in clusters are used. Values 
of   $n = 0.002 - 0.032$ nucleons/fm$^3$ and $T= 5 - 10$ MeV were obtained. 
Figure~\ref{fig:5}, taken from that work~\cite{R2013}, shows these results 
compared to results from several other 
analyses~\cite{Kowalski:2006ju,Qin,Zheng}. While the NSE model is applicable 
at only very low densities, below 
$n \approx 0.001$ fm$^{-3}$~\cite{Shlomo:2009},  the other models 
employed significantly higher densities and exhibit very reasonable 
agreement with each other. 
    
The comparison of results from   different techniques of extracting $T$ 
and $n_B$ from   experimental data are presented in Fig.~\ref{fig:5}. 
 Interestingly, the results of the coalescence model analysis~\cite{Qin}  
and the quantum fluctuation analysis~\cite{Zheng}  presented in 
Fig.~\ref{fig:5} lead to very similar results. Both produce results quite 
similar to the densitometer analysis based on calculated QS model 
equilibrium constants. The fact that the two different experimental 
results for the temperature and density regions explored are consistent 
with each other despite the fact that they are obtained from quite 
different beam energies, emitting sources  and analyses, suggests 
that an underlying unifying feature of the EOS is responsible. 
Indeed, further analysis by Mabiala {\it et al.}~\cite{20,21,Zheng} 
indicate that the data are sampling the vapor branch of the liquid 
gas coexistence curve and may be employed to determine the critical 
temperatures of mesoscopic nuclear systems,  within the framework of 
the Guggenheim systematics, in a manner analogous to previous 
treatments~\cite{Natowitz2,Elliott}. 

\subsection{Test of the nuclear matter EOS at low densities}

With our confidence in the temperature and density determinations bolstered 
by the consistency exhibited in Fig.~\ref{fig:5}, we have addressed various 
aspects of clustering in the low density nuclear matter produced. In 
theoretical models cluster mass fractions are commonly used to characterize 
the degree of clusterization in low density matter. However, different 
theoretical models include various different competing species. This 
leads to differences in particular mass fractions quoted.  If all 
relevant species are not included in the calculation, mass fractions 
cannot be accurately determined. To avoid this problem we have proposed 
that equilibrium constants for cluster formation be employed instead of 
mass fractions. In contrast to mass fractions, cluster formation 
equilibrium constants, such as that for $\alpha$ particle formation, i.e.,
\begin{equation}
                 K_c(\alpha)  =    \frac{n_\alpha}{n_n^2 n_p^2}     
\end{equation}
where $n_\alpha$, $n_n$, and $n_p$ are the $\alpha$ particle, neutron and 
proton densities, respectively, should be independent of proton fraction 
and choice of competing species.

\begin{figure}
\resizebox{0.53\textwidth}{!}{%
  \includegraphics{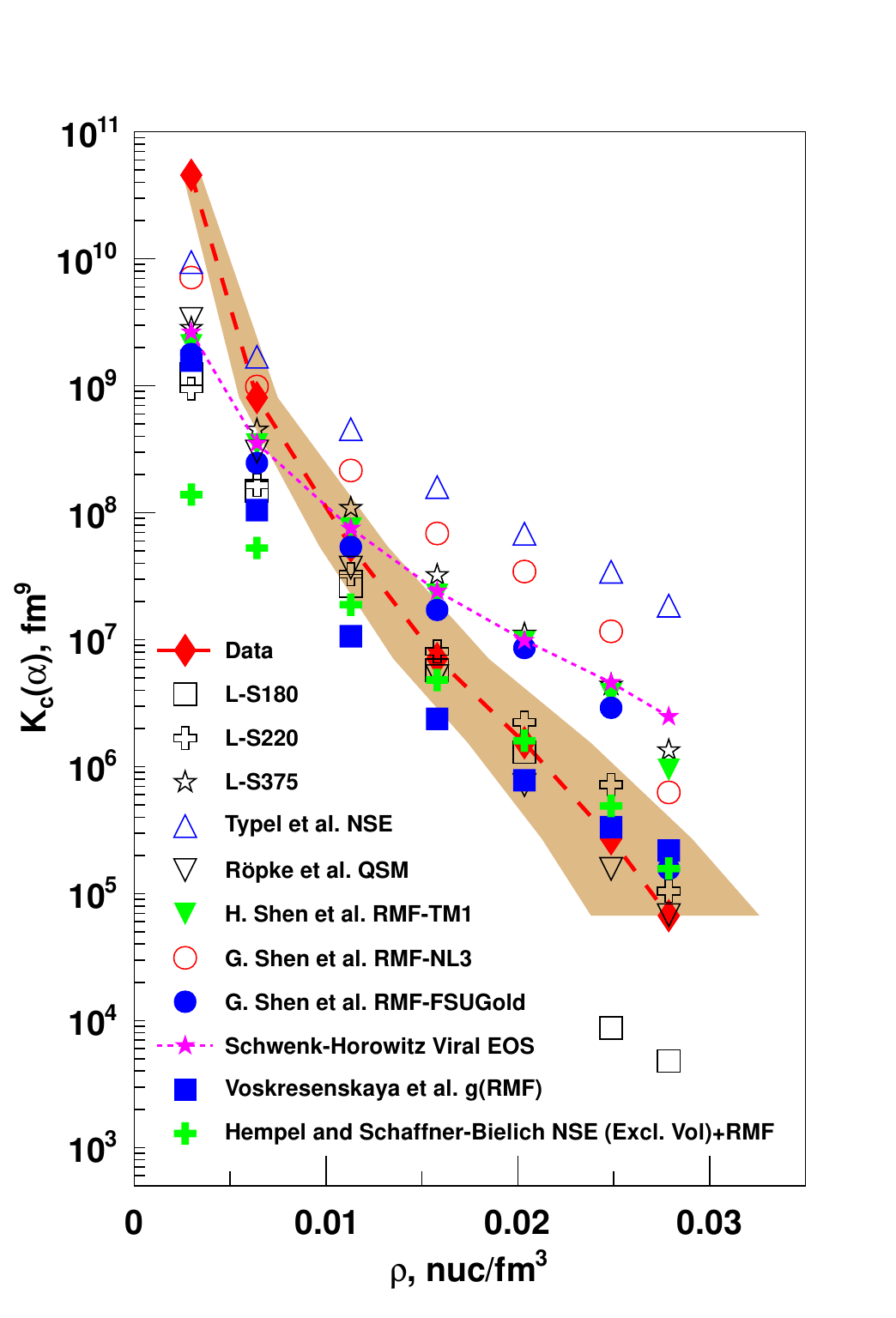}
}
\caption{ Comparison of experimental values of $K_c(\alpha)$ with those 
from various EOS calculations~\cite{Qin}.}
\label{fig:6}       
\end{figure}

Figure~\ref{fig:6} from reference~\cite{Qin}  depicts a comparison between 
our experimentally derived equilibrium constants and those resulting from 
models employing a variety of equations of state proposed for astrophysical 
applications \cite{Lattimer:1991nc,Shen:1998gq,Shen:2011,Hempel:2010,Typel:2009sy}.  
             
Not surprisingly the calculated values of the equilibrium constant tend to 
converge at the lower densities.   Even at the lowest densities sampled, 
however, there are significant differences.  At higher densities, 
0.01 to 0.03 nuc/fm$^3$, the various interactions employed all lead to 
a decrease of $K_c$ below that of the Nuclear Statistical Equilibrium (NSE) 
values of Typel {\it et al.}~\cite{Typel:2009sy}, as expected. However most 
of them lead to higher values of $K_c(\alpha)$ than those derived from the 
experiment.  The Lattimer-Swesty model~\cite{Lattimer:1991nc}  using Skyrme 
models with incompressibilities of 180 and 220 MeV and employing an excluded 
volume technique predict values slightly higher than the experimental values. 
This is also true for the Statistical Equilibrium model of Hempel and 
Schaffner-Bielich using an NL-3 interaction and also employing an excluded 
volume technique~\cite{Hempel:2010}.  The Quantum Statistical Model of 
R\"opke \textit{et al.}, which includes the  medium modifications of the 
cluster binding energies leads to values close to the experimental 
values~\cite{Ropke:2008qk,Ropke2011}.  The data provide important new 
constraints on the model calculations.    

\subsection{Shift of the binding energies and Mott points}

\begin{figure}
\resizebox{0.525\textwidth}{!}{%
  \includegraphics{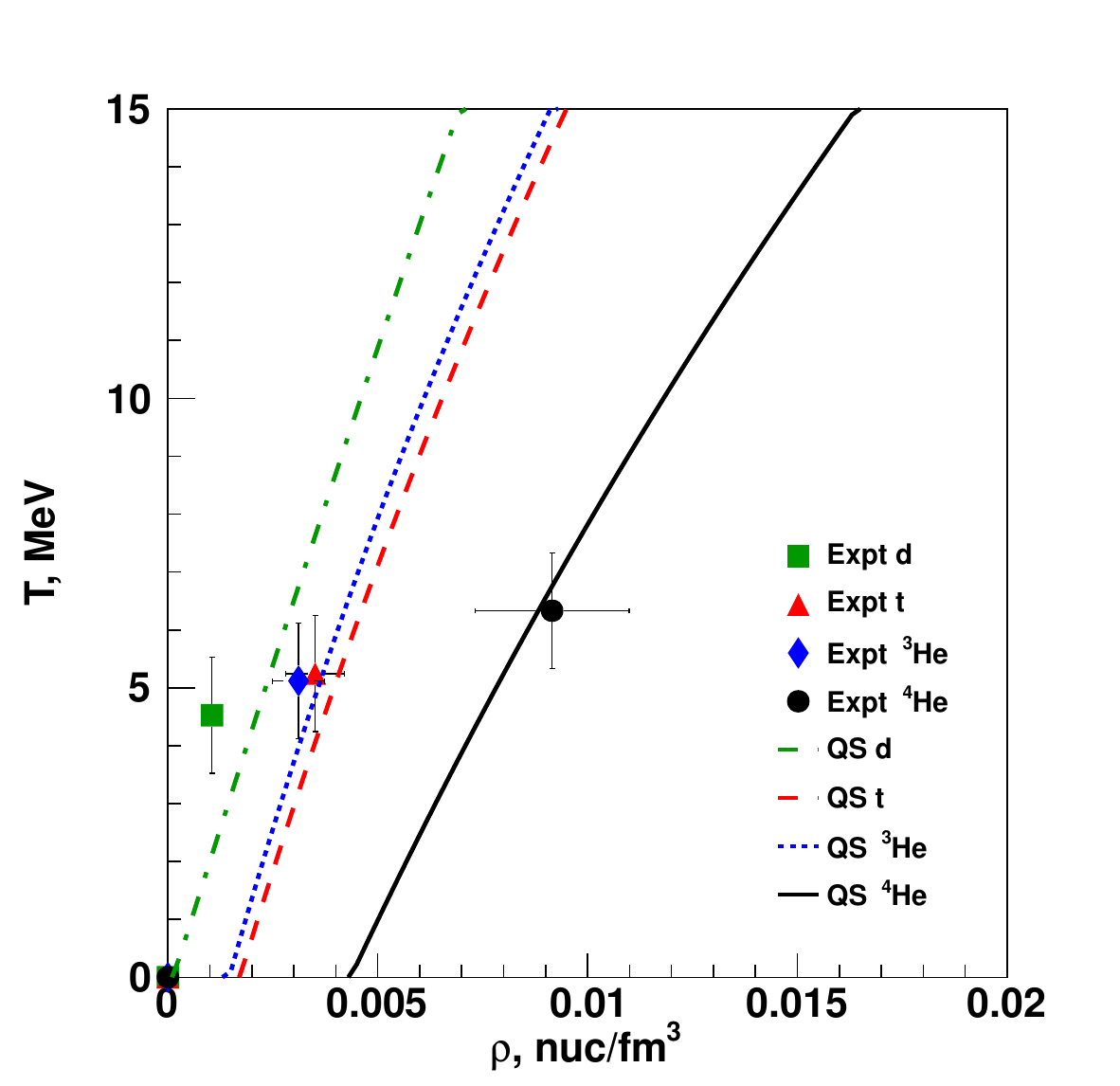}
}
\caption{ Comparison of experimentally derived Mott point densities and 
temperatures with theoretical values~\cite{Hagel}. Symbols represent the experimental 
data. Estimated errors on the temperatures are 10 \% and on the 
densities 20 \%. Lines show polynomial fits to the Mott  points presented 
in  reference~\cite{Typel:2009sy}.}
\label{fig:7}       
\end{figure}

Using the equilibrium constants it is possible to derive temperature and 
density dependent binding energies of $d, t, ^3$He and $\alpha$ clusters 
from the data~\cite{Hagel}. Since the observed temperatures and densities 
are correlated in our experiment the point at which an experimentally 
derived cluster binding energy is zero with respect to the surrounding 
medium corresponds to a particular combination of density and temperature. 
Thus, for our data, on clusters produced in collisions of 
47 $A$ MeV   $^{40}$Ar and  $^{64}$Zn projectiles with  $^{112}$Sn 
and $^{124}$Sn  target nuclei we are able to extract a single Mott point 
for each cluster. In Figure~\ref{fig:7} we present the values of the 
Mott temperatures and densities and compare them with the loci of the 
values of medium modified binding energies predicted by the thermodynamic 
Green's function method~\cite{Ropke:2008qk,Ropke2011}, see 
Typel {\it et al.}~\cite{Typel:2009sy}. This approach makes explicit use of 
an effective nucleon-nucleon interaction to account for medium effects on 
the cluster properties.  We see that the agreement between the predictions 
and the experimental results is quite good.

\section{Clusterization and the Symmetry Energy at Low Nucleon Density}

\subsection{Isoscaling and free symmetry energy}

To explore the symmetry energy in low density nuclear matter we first 
employed isoscaling techniques to derive symmetry free 
energies~\cite{MBTsang}. This analysis assumes that for two systems 
with similar temperatures but different $N/Z$ ratios, the ratio of yields 
of a particular isotope of mass number $A$, proton number $Z$, and neutron 
number $N$  in the two systems may be expressed as~\cite{tsang01,souza08} 
\begin{eqnarray}
\label{eq4}
\frac{Y_2(Z,N)}{Y_1(Z,N)}&=&
Ce^{((\mu_2(n)-\mu_1(n))N+(\mu_2(p)-\mu_1(p))Z)/T} \nonumber \\
&=&Ce^{\alpha N+\beta Z},
\end{eqnarray}
where $C$ is a constant and $\mu(n)$ and $\mu(p)$ are the neutron and 
proton chemical potentials.  The isoscaling parameters 
$\alpha=\big(\mu_2(n) - \mu_1(n)\big)/T$ and $\beta= \big(\mu_2(p) - \mu_1(p)\big)/T$, 
representing the difference in chemical potential between the two systems, 
may be extracted from suitable plots of yield ratios. Either parameter 
may then be related to the symmetry free energy, F$_{\rm sym}$. We take the $\alpha$ 
parameter, which is expected to be less sensitive to residual Coulomb 
effects.  With the usual convention that system 2 is richer in neutrons 
than system 1, 
\begin{equation}
\alpha=4F_{\rm sym}\left[\left(\frac{Z_1}{A_1}\right)^2-\left(\frac{Z_2}{A_2}\right)^2\right]/T\,,
\end{equation}
where $Z$ is the atomic number and $A$ is the mass number of the 
emitter~\cite{natowitz10,tsang01,souza08}. Thus, F$_{\rm sym}$ may be derived 
directly from determinations of system temperatures, $Z/A$ ratios, and 
isoscaling parameters. We emphasize that the present analysis is carried 
out for light species characteristic of the nuclear gas rather than, as 
in most previous analyses, for the intermediate mass fragments thought 
to be characteristic of the nuclear liquid. 

In this work we employ Eq.~(\ref{eq4}) with experimentally determined 
isoscaling parameters, $\alpha$, temperatures, $T$, and $Z/A$ ratios to 
determine the symmetry free energy coefficient F$_{\rm sym}$.  Assuming the 
quadratic behavior of the symmetry energy term in the mass formula we can write
$\alpha=4F_{\rm sym}\, \Delta(Z/A)^2/T$.  Here $\alpha$ is the isoscaling 
coefficient determined from yield ratios of $Z =1$ ejectiles of the two 
reactions, $F_{\rm sym}$ is the free symmetry energy and $ \Delta(Z/A)^2$ 
is the difference of the squared asymmetries of the sources in the two 
reactions. 

The isoscaling analysis has been employed (as a function of $v_{\rm surf}$) to 
determine $\alpha$ via the expression (\ref{eq4}).  With $ \Delta(Z/A)^2$ 
and the temperature determined from observed yields, the free symmetry 
energy is extracted~\cite{Kowalski:2006ju,Wada}  . 

\subsection{Internal symmetry energy}
From the free symmetry energy $F_{\rm sym}$  the internal symmetry 
energy $E_{\rm sym}$ can be derived if the symmetry entropy $S_{\rm sym}$ is 
known,
\begin{equation}
E_{\rm sym}=F_{\rm sym}+T S_{\rm sym}\,.
\end{equation}
Simple approximations for $S_{\rm sym}$ obtained from NSE or virial 
expansions, see~\cite{Kowalski:2006ju}, cannot be used to derive the 
symmetry energy from the free symmetry energy.  Given the general 
consistency of our data with the results of the QS calculations we have 
employed that model to determine the requisite symmetry entropies as a 
function of density and temperature. In contrast to the mixing entropy 
that leads to a larger entropy for uncorrelated symmetric matter in 
comparison with pure neutron matter, the formation of correlations, in 
particular clusters, will reduce the entropy in symmetric matter, see 
Fig.~\ref{fig:9} of Ref.~\cite{Typel:2009sy}.  For parameter values for which 
the yields of free nucleons in symmetric matter are small, the symmetry 
entropy may become positive for low temperatures.  The fraction of nucleons 
bound in clusters can decrease, e.g. due to increasing temperature or the 
dissolution of bound states at high densities due to the Pauli blocking. 
Then, the symmetric matter recovers its larger entropy so that the symmetry 
entropy becomes negative.

Within this analysis we have employed the calculated symmetry entropy 
coefficients from the QS model~\cite{Wada,natowitz10} to determine the 
entropy contribution to the symmetry free energy and extract values of the 
symmetry energy coefficients from the data. The symmetry energies derived 
in this manner are presented in Fig.~\ref{fig:8}. 

\subsection{Results for the symmetry energy and discussion}
\begin{figure}
\vspace {-0.5truein}
\hspace {-0.85truein}
\resizebox{0.75\textwidth}{!}{%
  \includegraphics{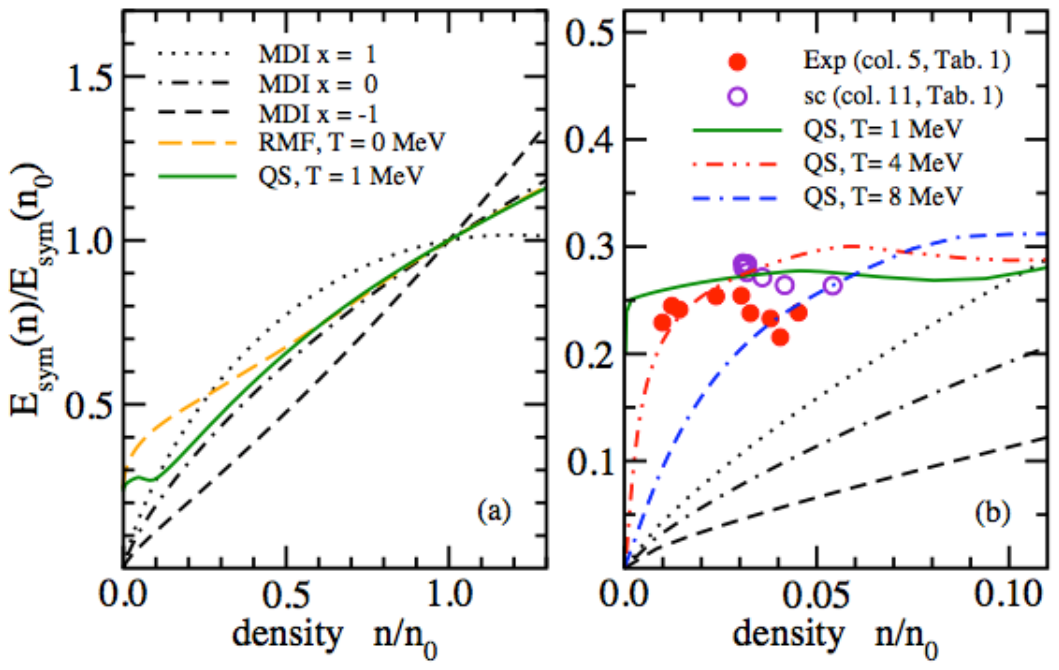}
}
\vspace {-0.9truein}
\caption{
Comparisons of the scaled internal symmetry energy 
$E_{\rm sym}(n)/E_{\rm sym}(n_0)$ 
as a function of the scaled total density $n/n_0$ for different approaches 
and the experiment.  (left panel):  The commonly used MDI parametrization of 
Chen \textit{et al.}~\cite{Che05} for $T=0$ and different asy-stiffnesses, 
controlled by the parameter $x$ (dotted, dot-dashed and dashed (black)lines). 
The result of the QS calculation including light clusters for temperature 
$T=1$~MeV is shown as the solid (green)line.  We also show the symmetry 
energy for a RMF calculation at $T=0$ where heavy clusters have been 
included (long-dashed (beige) line).  (right panel):  The internal scaled 
symmetry energy in an expanded low-density region.  Shown are again the MDI 
curves and the QS results for $T=4$ and 8~MeV.  We compare these with the 
experimental data with the NSE entropy (solid circles) and the results of 
the self-consistent calculation (open circles) from Ref.~\cite{natowitz10}. 
Recall that both $T$ and $n$ vary for the experimental data.
}
\label{fig:8}       
\end{figure}

The derived symmetry energy coefficients are plotted against density in 
Figure~\ref{fig:8} where they are compared to those which are predicted by  
the RMF and QS  model calculations~\cite{Typel2013}.  Note that the 
underlying RMF model for the quasiparticle description with 
$n_0=0.149$~fm$^{-3}$, $E_{\rm sym}(n_0)=32.73$~MeV gives a reasonable 
behavior at high density. However, at the lowest densities sampled the 
quasiparticle mean-field approach (RMF without clusters) disagrees strongly 
with the experimentally deduced symmetry energy while the QS approach 
gives a rather good agreement with the experimental data. This reflects 
the formation of clusters, primarily  $\alpha$ particles, not included in 
such calculations.  The model with medium effects describes quite well the 
low-density symmetry energy data that were extracted from our analysis of 
heavy-ion collisions.    

  \vspace{0.5cm}
\begin{figure}
\vspace {-0.5truein}
\resizebox{0.525\textwidth}{!}{%
   \includegraphics{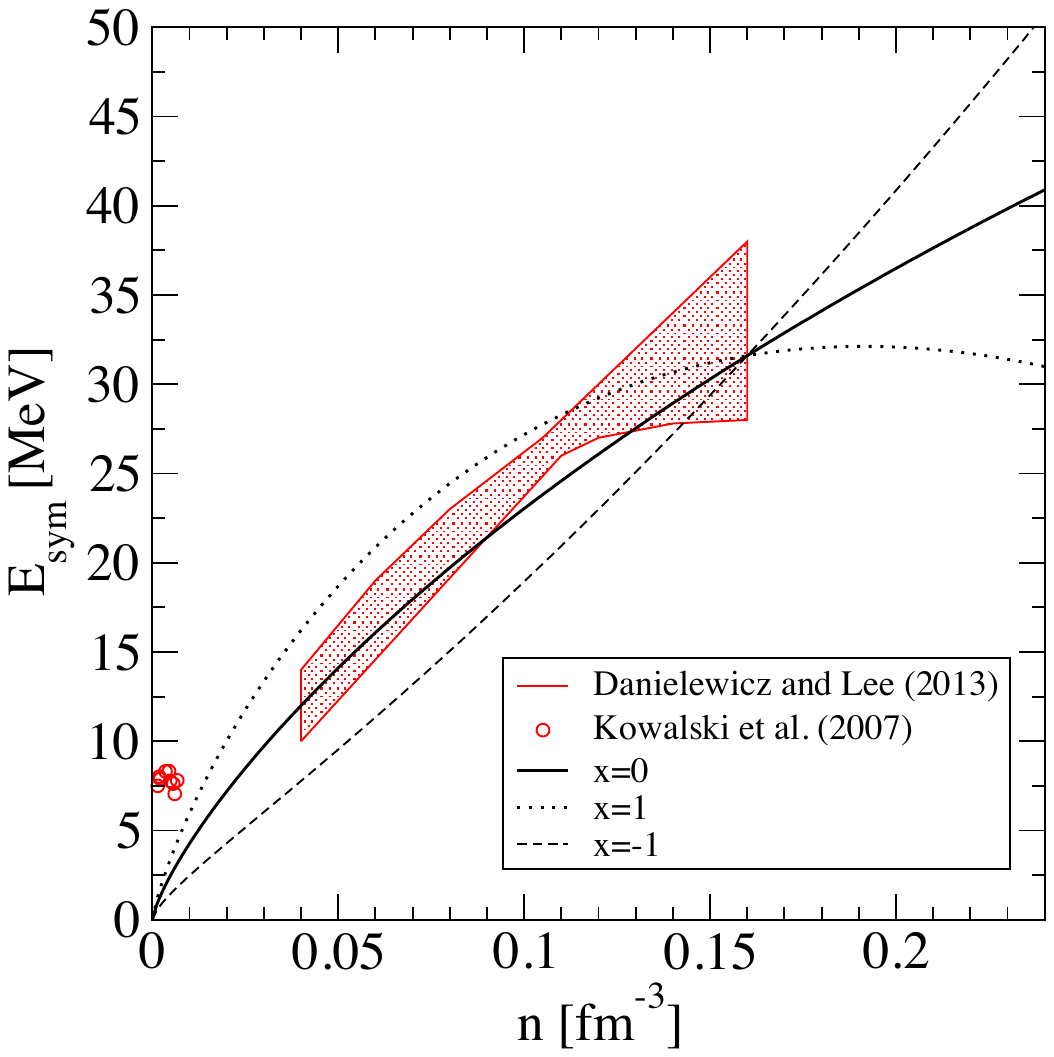}
\hspace {-1truein}
  \includegraphics{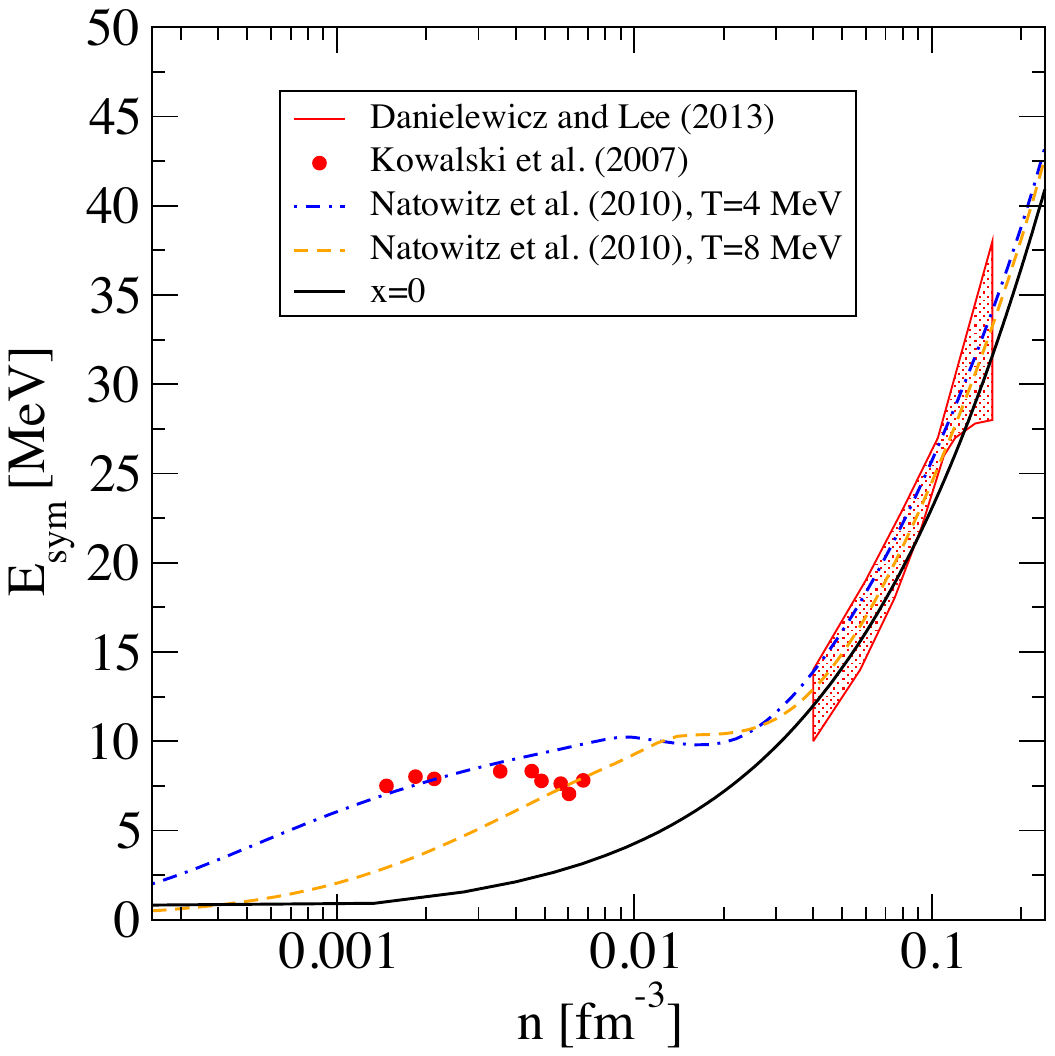}
}
\vspace {-0.25truein}
\caption{
Symmetry energy as function of density, linear scale (left panel) and
logarithmic scale (right panel). Experiment~\cite{Kowalski:2006ju} compared 
with predictions according to the MDI parametrization of 
Chen \textit{et al.}~\cite{Che05} with different parameter values $x$
and the QS model \cite{natowitz10}.
The hatched area shows constraints obtained from Isobaric Analog 
States (IAS) of Danielewicz and Lee~\cite{danielewicz13}.
Courtesy of David Blaschke.
}
\label{fig:8a}       
\end{figure}

In Fig.~\ref{fig:8} we compare the internal symmetry energy in different 
approaches with the experimental values.  In the left panel of the figure 
we show theoretical results for $T$ at or close to zero.  A widely used 
momentum-dependent parametrization of the symmetry energy (MDI) at 
temperature $T=0$~MeV was given in Refs.~\cite{Li:2008gp,Che05} and is 
shown by dotted, dot-dashed and dashed lines corresponding to different
values of the stiffness, governed by the parameter $x$.  However, all these 
parametrizations vanish in the low-density limit. We compare this to the QS 
result at $T=1$~MeV (since the QS approach is not suitable for constructing 
the correlated ground state at $T=0$).  In this approach the symmetry energy is finite at low density.
(Because of the finite temperature at extremely low densities 
of the order of $10^{-5}~{\rm fm}^{-3}$ this curve will also approach zero.)
Also note that the underlying RMF model for the quasiparticle description 
gives a reasonable behavior at high density similar to the MDI, $x$=0 
parametrization.  We thus see that the QS approach successfully interpolates 
between the clustering phenomena at low density and a realistic description 
around normal density.  The generalized RMF, $T=0$ curve 
(see~\cite{Typel:2009sy})  was constructed taking into account cluster 
formation, but demands further discussion with respect to the treatment 
of Coulomb interaction and phase transitions, see~\cite{Typel2013}.

Fig.~\ref{fig:8a} shows a comparison of the symmetry energy extracted from 
experimental data~\cite{Kowalski:2006ju} to predictions of Danielewicz and 
Lee~\cite{danielewicz13} and Chen {\it et al.}~\cite{Che05}.  The hatched area in the 
figure shows constraints obtained from analysis of isobaric analog
states~\cite{danielewicz13}.  Calculations in~\cite{natowitz10} combined
with this analysis suggest a smooth transition from low density to normal
nuclear density.

\begin{figure}
\resizebox{0.545\textwidth}{!}{%
  \includegraphics{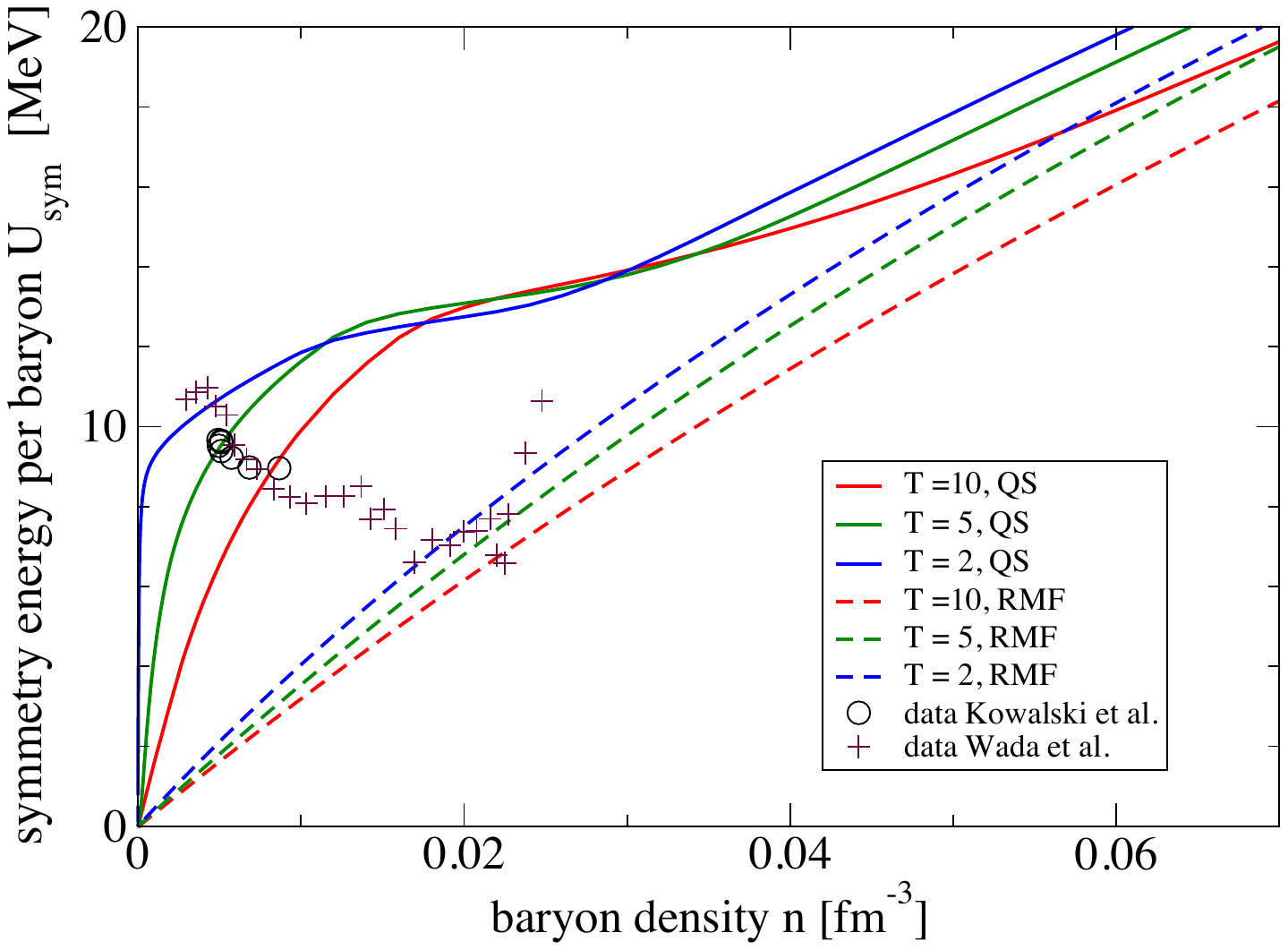}
}
\caption{
Internal symmetry energy coefficients as a function of baryon 
density in nucleons/fm$^3$. Experimental results~\cite{Kowalski:2006ju,Wada} are compared to results 
of QS model calculations, see Fig.~\ref{fig:8}. For the experimental data, 
$T$ varies in the region 3 - 11 MeV, see~\cite{Wada,natowitz10}.
}
\label{fig:9}       
\end{figure}

In the right panel of Fig.~\ref{fig:8} as well as in Fig.~\ref{fig:9} we 
compare the internal symmetry energy derived from the experimental 
data~\cite{Kowalski:2006ju,Wada,natowitz10} in an expanded low-density region with the RMF and 
QS results.  Again, it is clearly seen that the quasiparticle 
mean-field approach (RMF without clusters) disagrees strongly with the 
experimentally deduced symmetry energy while the QS approach gives a rather 
good agreement with the experimental data. 
 
We find that medium-dependent cluster formation has to be considered
in theoretical models to obtain the low-density dependence of the symmetry 
energy that is observed in experiments.  The frequently used presentation
for the density dependence of the symmetry energy that show vanishing 
symmetry energy at zero density and an increase that is linear with the density
is not correct because correlations and cluster formation are essential in the 
low-density region.  Note the strong dependence on temperature that is also 
not considered in the standard presentations, but has to be taken into account 
for applications in astrophysics and heavy ion collisions where temperatures 
in the range of several MeV are of interest.

\section{Conclusion and outlook}

In conclusion, our comparison of experimental data with results of model 
calculations strongly indicates that accurate modeling of clusterization 
in low density matter is critical for both nuclear and astrophysical 
applications.  A quantum-statistical model of nuclear matter, that includes 
the formation of clusters at densities below nuclear saturation, describes 
quite well the experimental low-density symmetry energy which was extracted 
from the analysis of heavy-ion collisions. 
 
Using the QS approach, the composition and the thermodynamic quantities of 
nuclear matter can be modeled over a large region of densities, temperatures 
and asymmetries.  In particular, it reproduces the statistical models of a 
gas consisting of various nuclei at low densities and the mean-field 
approaches that are applicable near the saturation density.  An important 
ingredient is the disappearance of  bound states at a certain density 
(denoted as Mott density) due to Pauli blocking.
 
Analyzing heavy ion collisions using the NIMROD multi-detector at the 
Cyclotron Institute at Texas A \& M University the medium modification 
of light fragments that leads to the dissolution has been shown.  Yields of 
light particles produced in the collisions of 
47 $A$ MeV $^{40}$Ar with $^{112}$Sn ,$^{124}$Sn and 
$^{64}$Zn with $^{112}$Sn, $^{124}$Sn were employed in thermal 
coalescence model analyses to derive densities and temperatures of the 
evolving emitting systems. The nuclear matter equation of state has been
tested, and significant deviations from the NSE have been found. 
The relevance of medium modifications that are obtained from the QS approach
have been proven by experiments.

Isoscaling analyses were used to determine the free symmetry energies of 
these systems. Comparisons of the experimental values with those of 
calculations made using a model which incorporates medium modifications 
of cluster binding energies reveal a very good agreement. The model 
calculated symmetry entropies have been used together with the experimental 
free symmetry energies to derive symmetry energies of nuclear matter at 
densities of $0.03 \leq n/n_0 \leq 0.2$  and temperatures in the range 
5 to 11 MeV.  It has been shown that the occurrence of bound states, 
absent in a mean-field approach, leads to a significant increase of the 
symmetry energy in the low-density region in contrast to a linear increase 
of the symmetry energy with density as predicted by many mean-field 
motivated models. In the low-density region, the symmetry energy is 
strongly depending on temperature.

{\bf Acknowledgement}:
This research was supported by CompStar, a Research Networking 
Programme of the European Science Foundation, and  in particular by grant No. DE-FG03-93ER40773 
(Texas A \& M) and by Robert A. Welch Foundation grant No. A0330 (JBN).

\end{document}